\documentclass[12pt]{iopart}

\usepackage{graphicx}
\usepackage{bm}
\usepackage{amssymb}     
\usepackage{epsfig}
\usepackage{pstricks,pst-grad}
\usepackage{bbm}

\begin{document}

\title{The $T=0$ random-field Ising model on a Bethe lattice with large coordination number: hysteresis and metastable states}

\date{\today}

\author{M.L.~Rosinberg}
%
\address{Laboratoire de Physique Th\'eorique de la Mati\`ere Condens\'ee, CNRS-UMR 7600, Universit\'e Pierre et Marie Curie, 4 place Jussieu, 75252 Paris Cedex 05, France}

\author{G.~Tarjus}
\address{Laboratoire de Physique Th\'eorique de la Mati\`ere Condens\'ee, CNRS-UMR 7600, Universit\'e Pierre et Marie Curie, 4 place Jussieu, 75252 Paris Cedex 05, France}

\author{F.J.~P\'erez-Reche}
%
\address{Department of Chemistry, University of Cambridge, Cambridge, CB2 1EW, UK}

\begin{abstract}

In order to elucidate  the relationship between rate-independent hysteresis and metastability in disordered systems driven by an external field, we study the Gaussian RFIM at $T=0$ on regular random graphs (Bethe lattice) of finite connectivity $z$ and compute to $O(1/z)$ (i.e. beyond mean-field) the quenched complexity associated with the one-spin-flip stable states with  magnetization $m$ as a function of the magnetic field $H$. When the saturation hysteresis loop  is smooth in the thermodynamic limit, we find that it coincides with the envelope of the typical metastable states (the quenched complexity vanishes exactly along the loop and is positive everywhere inside). On the other hand, the occurence of a jump discontinuity in the loop (associated with an infinite avalanche) can be traced back to the existence of a gap in the magnetization of the metastable states for a range of applied field, and the envelope of the typical metastable states is then reentrant. These findings confirm and complete earlier analytical and numerical studies.
\end{abstract}

\maketitle

\def\be{\begin{equation}}
\def\ee{\end{equation}}
\def\bea{\begin{eqnarray}}
\def\eea{\end{eqnarray}}

\section{Introduction}

At low temperature, disordered systems exhibit a large number of metastable states which are responsible for their slow relaxation dynamics and their irreversible, hysteretic response to an external field. In many cases this  response is very well reproducible and independent of the rate at which the field is changed, provided it is low enough. This shows that these systems always explore the same set of metastable states when driven by the external field and that thermally activated processes are irrelevant on  experimental time scales. Classical examples of such behavior are magnetic materials. When the external magnetic field is slowly cycled from large negative to large positive values and back, the induced magnetization displays a (saturation) hysteresis loop with a characteristic shape  that depends on the mechanisms for the formation and evolution of the  magnetic domains\cite{B1998}.  Other field histories give rise to a complicated subloop structure that is also  well reproducible.

By definition, the future evolution of a hysteretic system depends on its past history, and the theory of rate-independent hysteresis generally focuses on a dynamical description in order to describe history-dependent effects such as return-point memory\cite{B1998,S1993,KZ2007}. However, the  hysteresis loop may also be viewed as a special path among the metastable states and one may wonder whether it is possible to characterize this path without having to follow the dynamical evolution. Surprisingly, a general answer to this question is not  yet available. 
In the case of attractive (ferromagnetic) interactions, however, an important fact is   known: if the dynamics obeys the so-called no-passing rule\cite{S1993} (i.e. preserves some natural partial ordering of the microscopic configurations), there are no metastable states outside the saturation hysteresis loop\cite{DRT2005}. Since, as a general rule, the number of metastable states scales exponentially with the system size, this result implies that the corresponding entropy density (or complexity) is not positive outside the loop. What still remains to be understood is the situation {\it inside} the loop.

A convenient framework for studying this problem is the zero-temperature Random Field Ising model (RFIM) driven by an external field. This is a model for disordered ferromagnets which is relevant to a large class of materials where disorder induces random fields\cite{SDP2006}. With the standard Glauber dynamics (which does satisfy the no-passing rule at $T=0$\cite{S1993}), the metastable states are the so-called single-spin-flip stable states, each state being characterized by its energy and magnetization. The  quantity to be studied in relation to hysteresis is ${\cal N}(m,H)$, the number of metastable states at the field $H$ with given magnetization per spin $m$. One can then reformulate the above question as follows: in which region of the field-magnetization plane is the logarithm of ${\cal N}(m,H)$ an extensive quantity ? In particular, does the saturation hysteresis loop identifies with the boundary of this region ? To answer these questions, one must study the properties of  a {\it typical} system and compute the {\it quenched} complexity $\Sigma_Q(m,H)$ of the metastable states\cite{DRT2005}. This is  a nontrivial analytical or numerical task in general, and there exists  very few  results on the typical number of metastable states in finite-connectivity disordered spin systems. For the Gaussian RFIM, the only analytical result so far has been obtained in one dimension\cite{DRT2005} (or, equivalently, on a Bethe lattice of connectivity $z=2$), showing that the curve $\Sigma_Q(m,H)=0$  coincides with the hysteresis loop. Numerical results on the Bethe lattice of coordination $z=4$ and on the cubic lattice\cite{PRT2008} strongly suggest that this property always holds when the hysteresis loop is smooth in the thermodynamic limit (which occurs when the disorder is large enough). This means that the typical number of metastable states is exponentially large everywhere inside the loop. The situation is more complicated when the  loop has a jump at some critical value of the field, which corresponds to an out-of-equilibrium phase transition associated with the appearance of a macroscopic ``avalanche''\cite{S1993}. This happens at low disorder in three and higher dimensions\cite{S1993,DS1996} or on a Bethe lattice for $z\ge 4$\cite{DSS1997}. It has been suggested\cite{DRT2005} that this jump reveals the existence of a gap in the magnetization of the typical metastable states in a {\it finite} range of $H$, gap which may also explain the reentrant hysteresis loops observed when changing the driving mechanism\cite{IRV2006}. This assumption is also supported by numerical computations\cite{PRT2008}, but again there is  no analytical proof. The aim of the present work is to give further evidence to this scenario  (both above and below the critical disorder) by computing analytically the quenched complexity of the Gaussian RFIM on a Bethe lattice to order $1/z$, and comparing with the exact expression of the hysteresis loop\cite{DSS1997} at the same order. This will also help us to clarify the approach to the mean-field fully-connected limit\cite{RTP2008}.

The paper is organized as follows. In section II, we define the model and present the general formalism for calculating the quenched complexity. Section III is devoted to the mean-field limit which is recovered when $z\rightarrow\infty$. The calculation of the complexity at the order $1/z$ is presented in section IV and the numerical results are discussed in section V.  We finish with a brief discussion and some perspectives in section VI. More details on the analytical calculations are given in the appendices.

\section{General formalism}

We start with  the RFIM Hamiltonian
\begin{equation}
\label{Eq0} 
{\cal H}=-J\sum_{<i,j>}s_i s_j-\sum_i (h_i+H)s_i
\end{equation}
describing  a collection of $N$ Ising spins ($s_i=\pm 1$) placed on the vertices of a graph of connectivity $z$ and submitted to an external uniform field $H$.  A ferromagnetic interaction $J>0$ couples the spin $i$ to its $z$ neighbours $j$ and $\{h_i\}$ is a collection of random fields drawn identically and independently from a Gaussian probability distribution ${\cal P}(h)$ with zero mean and standard deviation $\Delta$. In the  zero-temperature system with metastable dynamics\cite{S1993}, the field $H$ is adiabatically varied (for instance from $-\infty$ to $+\infty$ and back) and each spin is aligned with its local effective field
\begin{equation}
\label{Eq2} f_i=J\sum_{j/i}s_j+h_i+H 
\end{equation}
so that
\begin{equation}
\label{Eq1} s_i=\mbox{sgn}(f_i) \ .
\end{equation}
In the thermodynamic limit, the resulting hysteresis loop $m(H)$ (where $m=(1/N)\sum_i s_i$ is the magnetization per spin) is smooth for $\Delta>\Delta_c(z)$ and discontinuous for $\Delta<\Delta_c(z)$, where $\Delta_c(z)$ is a critical value of the disorder at which avalanches of all sizes occur (an avalanche corresponds to a jump in the magnetization curve $m(H)$). A complete analytical description of this behavior is available in the mean-field model\cite{S1993,DS1996} which can be obtained for instance by placing the spins on a fully-connected lattice. In this limit, there is no hysteresis above $\Delta_c^0=\lim_{z\rightarrow \infty}\Delta_c(z)=J\sqrt{2/\pi}$ and the number of metastable states is not exponentially large\cite{RTP2008} (but it may be larger than $1$, even along the so-called ``unstable'' branch for $\Delta<\Delta_c^0$). 

Our goal in this paper is to compute the first non-zero term in the $1/z$ expansion of the complexity $\Sigma_Q(m,H)$.  Although the corrections to the mean-field result are certainly different on the Bethe lattice and on the hypercubic lattice, we only consider the former case since an analytical expression of the hysteresis curve is available for any $z$\cite{DSS1997}. This also allows us to use the machinery developped in Ref.\cite{DRT2005} to compute $\Sigma_Q(m,H)$. The possible influence of geometry will be briefly discussed in the conclusion.

 $\Sigma_Q(m,H)$ is defined as
\begin{eqnarray}
\label{Eq11}
\Sigma_Q(m,H)=\lim_{N\to \infty}\frac{1}{N} \overline{\ln {\cal N}(m,H)}=\lim_{n\to 0} \frac{1}{n} \lim_{N\to \infty}\frac{1}{N}  \Big[\overline{{\cal N}(m,H)^n}-1\Big]
\end{eqnarray}
where, as usual, the order of the limits $N\to \infty$ and the number of replicas $n\to 0$ has been inverted. The starting point of our calculation is the following expression of $\overline{{\cal N}(m,H)^n}$ in the large-$N$ limit\cite{DRT2005}:
\begin{eqnarray}
\label{Eq12}
\overline{{\cal N}(m,H)^n}&\sim \int \prod_{a=1}^n dg^a \prod_{\mbox{\boldmath $\sigma$},\mbox{\boldmath $\tau$}} dc(\mbox{\boldmath $\sigma$},\mbox{\boldmath $\tau$})e^{N {\cal F}(\{c\},\mbox{\boldmath $g$})}
\end{eqnarray}
where
\begin{eqnarray}
\label{Eq12a}
{\cal F}(\{c\},\mbox{\boldmath $g$})=\Lambda(\{c\},\mbox{\boldmath $g$})+ \frac{z}{2}-\frac{z}{2}\sum_{\mbox{\boldmath $\sigma$},\mbox{\boldmath $\tau$}}c(\mbox{\boldmath $\sigma$},\mbox{\boldmath $\tau$})c(\mbox{\boldmath $\tau$},\mbox{\boldmath $\sigma$}) - m \sum_{a=1}^n g^a
\end{eqnarray}
and
\begin{eqnarray}
\label{Eq13}
\fl \Lambda(\{c\},\mbox{\boldmath $g$})=\ln \sum_{\mbox{\boldmath $\sigma$}}\int {\cal P}(h) dh\int d {\bf x} d {\bf y} \ e^{i{\bf y}.({\bf x}-{\bf H}-{\bf h})}\Big[\sum_{\mbox{\boldmath $\tau$}}c(\mbox{\boldmath $\tau$},\mbox{\boldmath $\sigma$})e^{-iJ{\bf y}.\mbox{\boldmath $\tau$}}\Big]^z  \prod_{a=1}^n e^{ g^a\sigma^a}\Theta(x^a \sigma^a) \nonumber\\
\end{eqnarray}
where $\Theta(x)$ is the Heaviside function. In these equations, the bold symbols denote $n$-component vectors in replica space, e.g. ${\bf x}=\{x^a; a=1,...n\},$ (with ${\bf h}=h{\bf 1}$ and  ${\bf H}=H{\bf 1}$); $c(\mbox{\boldmath $\sigma$},\mbox{\boldmath $\tau$})$ is an order parameter, function of the two binary vectors $\mbox{\boldmath $\sigma$}$ and $\mbox{\boldmath $\tau$}$ ($\sigma^a=\pm1$, $\tau^a=\pm1$), which allows the decoupling of  the sums over the sites $i$. It generalizes the single-argument order-parameter function $c(\mbox{\boldmath $\sigma$})$  introduced in Ref. \cite{M1998} to study finite-connectivity spin glasses (see also Ref.\cite{BS2002}). ${\bf g}=\{g^a\}$ is a set of Lagrange multipliers associated with the constraint  $\sum_is_i^a=Nm$ in each replica. (The interested reader is referred to Ref. \cite{DRT2005} for the derivation of these equations\cite{note1}. Apart from these  three equations, the present paper is self-contained.)

In the large-$N$ limit, the integral in Eq. (\ref{Eq12}) can be evaluated by the method of steepest descent. The  $2^{2n}$ order parameters $c(\mbox{\boldmath $\sigma$},\mbox{\boldmath $\tau$})$ and the $n$ Lagrange mutipliers $g^a$ are  determined through the saddle-point equations that correspond to the extremization of the functional {\cal F}(\{c\},\mbox{\boldmath $g$}),
\be
\label{Eq14}
c(\mbox{\boldmath $\sigma$},\mbox{\boldmath $\tau$})=\frac{1}{z}\ \frac{\partial \Lambda(\{c\},\mbox{\boldmath $g$})}{\partial c(\mbox{\boldmath $\tau$},\mbox{\boldmath $\sigma$})} 
\ee
\begin{eqnarray}
\label{Eq14a}m= \frac{\partial \Lambda(\{c\},\mbox{\boldmath $g$})}{\partial g^a} \ ,
\end{eqnarray}
where $a$ is any of the replicas $1,...,n$.  One can  easily see that the solution of Eq. (\ref{Eq14}) (that will be denoted  as $c^*(\mbox{\boldmath $\sigma$},\mbox{\boldmath $\tau$})$, dropping the dependence on $\mbox{\boldmath $g$}$ and $H$) satisfies the normalization condition 
\be
\label{Eq15}
\sum_{\mbox{\boldmath $\sigma$},\mbox{\boldmath $\tau$}}c^*(\mbox{\boldmath $\sigma$},\mbox{\boldmath $\tau$})c^*(\mbox{\boldmath $\tau$},\mbox{\boldmath $\sigma$})=1 
\ee
so that  Eqs. (\ref{Eq14a}) can be rewritten as\cite{DRT2005}
\begin{eqnarray}
\label{Eq16}
m=\sum_{\mbox{\boldmath $\sigma$},\mbox{\boldmath $\tau$}}
\sigma^a c^*(\mbox{\boldmath $\sigma$},\mbox{\boldmath $\tau$})c^*(\mbox{\boldmath $\tau$},\mbox{\boldmath $\sigma$}) \ .
\end{eqnarray}

Assuming that the solution of Eq. (\ref{Eq16}) in the limit $n\rightarrow 0$ is the same for all replicas, namely $g^{*a}=g^*(m,H)$, we  expand $\Lambda^{*}(m,H)\equiv \Lambda(c^*(\mbox{\boldmath $\sigma$},\mbox{\boldmath $\tau$}),g^*)$ in powers of $n$, 
\be
\Lambda^{*}(m,H)=\Lambda^{*(0)}(m,H)+ n\Lambda^{*(1)}(m,H) +O(n^2) \ ,
\ee
which, after insertion in Eq.(\ref{Eq12}), yields
\begin{equation}
\label{Eq17}
\Sigma_Q(m,H)=\Lambda^{*(1)}(m,H)-mg^*(m,H) \ ,
\end{equation}
where we have used that $\Lambda^{*(0)}(m,H)$ is equal to zero, which is necessary to obtain a well-defined $n\rightarrow 0$ limit.

As discussed in Refs. \cite{DRT2005,PRT2008}, it is in fact  convenient to treat the Lagrange multiplier $g$ as a free parameter and to consider that the magnetization is a function of $g$ given by the solution of Eq. (\ref{Eq16}) in the limit $n\rightarrow 0$. Defining  $\Lambda^{*}(g,H)\equiv \Lambda(c^*(\mbox{\boldmath $\sigma$},\mbox{\boldmath $\tau$}),g)$, we then write
\begin{equation}
\label{Eq17a}
\Sigma_Q(g,H)\equiv \Sigma_Q(m^*(g,H),H)=\Lambda^{*(1)}(g,H)-m^*(g,H)g
\end{equation}
where $m^*(g,H)=\lim_{n\rightarrow 0}\sum_{\mbox{\boldmath $\sigma$},\mbox{\boldmath $\tau$}}
\sigma^a c^*(\mbox{\boldmath $\sigma$},\mbox{\boldmath $\tau$})c^*(\mbox{\boldmath $\tau$},\mbox{\boldmath $\sigma$}) =\partial\Lambda^{*(1)}(g,H)/ \partial g$. 
In other words, $\Sigma_Q(m,H)$ and $\Lambda^{*(1)}(g,H)$ are mutually connected by a Legendre transform with $-g^*(m,H)$ being the slope of the complexity curve, i.e. $g^*(m,H)=-\partial \Sigma_Q(m,H)/ \partial m$.  It is crucial that the mapping $\{m^{*}(g,H), \Sigma_Q(g,H)\} \mapsto  \Sigma_Q(m,H)$ is unambiguously defined, as will be discussed in section V. Note also that the total complexity at the field $H$, irrespective of the magnetization, is obtained by setting $g=0$ in Eq.(\ref{Eq17a}). This corresponds to the maximum  of $\Sigma_Q(m,H)$ that occurs at a magnetization $m^{max}(H)=m^*(g=0,H)$.

We now turn to the  $1/z$ expansion of $\Sigma_Q(m,H)$. Using the rescaling $Jz\rightarrow J$, we rewrite Eq. (\ref{Eq13}) as
\be
\label{Eq18}
\Lambda(\{c\},{\bf g})=\ln \sum_{\mbox{\boldmath $\sigma$}}e^{gs}\int {\cal P}(h) dh\int d {\bf x} d {\bf y} \ e^{\phi({\bf x},{\bf y})} \prod_a \Theta(x^a \sigma^a)
\ee
where $s=\sum_a\sigma^a$ and 
\be
\label{Eq19}\phi({\bf x},{\bf y})=i{\bf y}.({\bf x}-{\bf H}-{\bf h})+z\ln \sum_{\mbox{\boldmath $\tau$}}c(\mbox{\boldmath $\tau$},\mbox{\boldmath $\sigma$})e^{-i\frac{J}{z}{\bf y}.\mbox{\boldmath $\tau$}} \ .
\ee

We then  formally expand Eqs. (\ref{Eq14}) and (\ref{Eq15}) in powers of $1/z$, assuming that
\be
\label{Eq21}
\Lambda^*(g,H)=z\Lambda_{-1}^*(g,H)+\Lambda_0^*(g,H)+\frac{1}{z}\Lambda_1^*(g,H)+...
\ee
\be
\label{Eq21a}
m^*(g,H)=m_0^*(g,H)+\frac{1}{z}m_1^*(g,H)+...
\ee
and
\be
\label{Eq22}c^*(\mbox{\boldmath $\sigma$},\mbox{\boldmath $\tau$})=c_0^*(\mbox{\boldmath $\sigma$},\mbox{\boldmath $\tau$})+
\frac{1}{z} c_1^*(\mbox{\boldmath $\sigma$},\mbox{\boldmath $\tau$})+\frac{1}{z^2} c_2^*(\mbox{\boldmath $\sigma$},\mbox{\boldmath $\tau$})+...
\ee
This yields a set of coupled equations to be solved at each order in $1/z$. The main difficulty  of the calculation is that successive orders are not independent. Indeed, one has
\be
\label{Eq23}
\phi({\bf x},{\bf y})=z\ln \Big (\sum_{\mbox{\boldmath $\tau$}}c_0^*(\mbox{\boldmath $\tau$},\mbox{\boldmath $\sigma$})\Big)+\phi_0({\bf x},{\bf y})+\frac{1}{z} \phi_1({\bf x},{\bf y})+\frac{1}{z^2}\phi_2({\bf x},{\bf y})+...
\ee
with
\numparts
\label{Eq24}
\begin{eqnarray}
\fl\phi_0({\bf x},{\bf y})&=i{\bf y}.({\bf x}-{\bf H}-{\bf h})+\frac{\sum_{\mbox{\boldmath $\tau$}}[c_1^*(\mbox{\boldmath $\tau$},\mbox{\boldmath $\sigma$})-iJ{\bf y}.\mbox{\boldmath $\tau$} c_0^*(\mbox{\boldmath $\tau$},\mbox{\boldmath $\sigma$})]}{\sum_{\mbox{\boldmath $\tau$}}c_0^*(\mbox{\boldmath $\tau$},\mbox{\boldmath $\sigma$})}
\end{eqnarray}
\begin{eqnarray}
\fl\phi_1({\bf x},{\bf y})=\frac{\sum_{\mbox{\boldmath $\tau$}}[c_2^*(\mbox{\boldmath $\tau$},\mbox{\boldmath $\sigma$})-iJ{\bf y}.\mbox{\boldmath $\tau$} c_1^*(\mbox{\boldmath $\tau$},\mbox{\boldmath $\sigma$}) 
-\frac{J^2}{2}({\bf y}.\mbox{\boldmath $\tau$})^2c_0^*(\mbox{\boldmath $\tau$},\mbox{\boldmath $\sigma$})]}{\sum_{\mbox{\boldmath $\tau$}}c_0^*(\mbox{\boldmath $\tau$},\mbox{\boldmath $\sigma$})}
-\frac{1}{2}\Big[\frac{\sum_{\mbox{\boldmath $\tau$}}[c_1^*(\mbox{\boldmath $\tau$},\mbox{\boldmath $\sigma$})-iJ{\bf y}.\mbox{\boldmath $\tau$} c_0^*(\mbox{\boldmath $\tau$},\mbox{\boldmath $\sigma$})]}{\sum_{\mbox{\boldmath $\tau$}}c_0^*(\mbox{\boldmath $\tau$},\mbox{\boldmath $\sigma$})}\Big]^2\nonumber\\
\end{eqnarray}
\endnumparts
etc.... Inserting Eq. (\ref{Eq18}) in Eq. (\ref{Eq14}) and using these expansions, we obtain 
\be
\label{Eq25}
\Lambda_{-1}^*+\ln \sum_{\mbox{\boldmath $\tau$}}c_0^*(\mbox{\boldmath $\tau$},\mbox{\boldmath $\sigma$})=0 
\ee
 at  leading order, and
\be
\label{Eq26}
\fl c_0^*(\mbox{\boldmath $\sigma$},\mbox{\boldmath $\tau$})=e^{-\Lambda_0^*+gs}\int {\cal P}(h) dh\int d {\bf x} d {\bf y} \ e^{\phi_0({\bf x},{\bf y})} \prod_a \Theta(x^a \sigma^a)\ ,
\ee
\begin{eqnarray}
\label{Eq27}
\fl c_1^*(\mbox{\boldmath $\sigma$},\mbox{\boldmath $\tau$})=e^{-\Lambda_0^*+gs}\int {\cal P}(h) dh\int d {\bf x} d {\bf y} \ e^{\phi_0({\bf x},{\bf y})} \prod_a \Theta(x^a \sigma^a)\Big[-\Lambda_1^*+\phi_1({\bf x},{\bf y})-iJ{\bf y}.\mbox{\boldmath $\tau$} \nonumber\\
-\sum_{\mbox{\boldmath $\tau$}'}[c_1^*(\mbox{\boldmath $\tau$'},\mbox{\boldmath $\sigma$})-iJ{\bf y}.\mbox{\boldmath $\tau$}' c_0^*(\mbox{\boldmath $\tau$}',\mbox{\boldmath $\sigma$})\Big]
\end{eqnarray}
at the two next orders. Similarly, the expansion of the normalization equation, Eq. (\ref{Eq15}), yields
\be
\label{Eq29}
\sum_{\mbox{\boldmath $\sigma$},\mbox{\boldmath $\tau$}}c_0^*(\mbox{\boldmath $\sigma$},\mbox{\boldmath $\tau$})c_0^*(\mbox{\boldmath $\tau$},\mbox{\boldmath $\sigma$})=1 \ ,
\ee 
\be
\label{Eq30}
\sum_{\mbox{\boldmath $\sigma$},\mbox{\boldmath $\tau$}}c_0^*(\mbox{\boldmath $\sigma$},\mbox{\boldmath $\tau$})c_1^*(\mbox{\boldmath $\tau$},\mbox{\boldmath $\sigma$})=0 \ ,
\ee
etc...
(Note that the dependence on $g$ and/or $H$ is not always indicated in order to simplify the notations. We shall do the same in the following, except when needed explicitly.)

\section{Mean-field limit ($z\rightarrow \infty$)}

We first consider the limit $z\rightarrow \infty$ and check that the results of the mean-field model\cite{S1993,DS1996,RTP2008} are recovered.
The calculation, which in this setting is not trivial, proceeds as follows. Inserting Eq. (21a) into Eq. (\ref{Eq26}) and performing the integrations over ${\bf y}$ and ${\bf x}$, we  obtain
\be
\label{Eq32}
c_0^*(\mbox{\boldmath $\sigma$},\mbox{\boldmath $\tau$})=e^{-\Lambda_0^*+gs+\sum_{\mbox{\boldmath $\tau$}'}c_1^*(\mbox{\boldmath $\tau$}',\mbox{\boldmath $\sigma$})/\sum_{\mbox{\boldmath $\tau$}'}c_0^*(\mbox{\boldmath $\tau$}',\mbox{\boldmath $\sigma$})}\int dh {\cal P}(h)\prod_a \Theta(x^{*a} \sigma^a)
\ee
with
\be
\label{Eq33}x^{*a}(\mbox{\boldmath $\sigma$})=H+h+J\frac{\sum_{\mbox{\boldmath $\tau$}'}\tau'^{a}c_0^*(\mbox{\boldmath $\tau$'},\mbox{\boldmath $\sigma$})} {\sum_{\mbox{\boldmath $\tau$}'}c_0^*(\mbox{\boldmath $\tau$}',\mbox{\boldmath $\sigma$})}\ .
\ee
Since the r.h.s. of Eq. (\ref{Eq32}) does not depend on $\mbox{\boldmath $\tau$}$, $c_0^*$ is a function of  its first argument only that will be denoted as $c_0^*(\mbox{\boldmath $\sigma$})$. As a result, the quantity $x^{*a}$ does not depend on $\mbox{\boldmath $\sigma$}$. It also does not depend on $a$ if one assumes that all replicas are equivalent. Therefore, because of the integration over $h$, the only functions that differ from zero are $c_0^*(\mbox{\boldmath $1$})$ and $c_0^*(- \mbox{\boldmath $1$})$ (i.e. all $\sigma^a$ are equal to $1$ or $-1$). From  Eq. (\ref{Eq29}), these two functions satisfy
\be
\label{Eq34}
c_0^*(\mbox{\boldmath $1$})+c_0^*(- \mbox{\boldmath $1$})=1
\ee
(the other  solution of Eq. (\ref{Eq29}), $c_0^*(\mbox{\boldmath $1$})+c_0^*(- \mbox{\boldmath $1$})=-1$, is not physically relevant). Eq. (\ref{Eq34}) is actually a prerequisite for having a finite limit when $z\rightarrow\infty$,  since it implies from Eq. (\ref{Eq25}) that $\Lambda_{-1}^*=0$ in the expansion (\ref{Eq21}). Defining $X_1(\mbox{\boldmath $\sigma$})=\sum_{\mbox{\boldmath $\tau$}}c_1^*(\mbox{\boldmath
$\tau$},\mbox{\boldmath $\sigma$)}$    and $\Delta c_0^*=c_0^*(\mbox{\boldmath $1$})-c_0^*(- \mbox{\boldmath $1$})$,  we then rewrite Eq. (\ref{Eq32}) and (\ref{Eq34}) as
\be
\label{Eq35}
\frac{1}{2}[1\pm \Delta c_0^*]=e^{-\Lambda_0^* \pm gn+X_1(\pm\mbox{\boldmath $1$})}\int dh {\cal P}(h)\Theta(\pm x^*)
\ee
and
\be
\label{Eq35a}
e^{\Lambda_0^*}=e^{gn+X_1(\mbox{\boldmath $1$})}\int dh {\cal P}(h)\Theta( x^*)+e^{-gn+X_1(-\mbox{\boldmath $1$})}\int dh {\cal P}(h)\Theta(- x^*)
\ee
where
\be
\label{Eq36}
 x^*=H+h+J\Delta c_0^*\ .
\ee
This is not yet a closed set of equations for computing $\Delta c_0^*$ and  $\Lambda_0^* $  because the two quantities $X_1(\pm \mbox{\boldmath $1$})$  depend on  $c_1^*(\mbox{\boldmath $\sigma$},\mbox{\boldmath $\tau$)}$. Therefore, the equations at the next order in $1/z$, Eqs. (\ref{Eq27}) and (\ref{Eq30}), must be also considered. The latter can be written as
\be
\label{Eq36a}
(1+\Delta c_0^*)X_1(\mbox{\boldmath $1$})+(1-\Delta c_0^*)X_1(-\mbox{\boldmath $1$})=0
\ee
so that we only need to extract from Eq. (\ref{Eq27}) the quantity $\Delta X_1=X_1(\mbox{\boldmath $1$})-X_1(-\mbox{\boldmath $1$})$. A quick look at this equation shows that the solution takes the form 
\begin{eqnarray}
\label{Eq39}
c_1^*(\mbox{\boldmath $\sigma$},\mbox{\boldmath $\tau$})=F(\mbox{\boldmath $\sigma$})+e^{-\Lambda_0^* +gs+X_1(\mbox{\boldmath $\sigma$})}I(\mbox{\boldmath $\sigma$},\mbox{\boldmath $\tau$})
\end{eqnarray}
where 
\begin{eqnarray}
\label{Eq40}
I(\mbox{\boldmath $\sigma$},\mbox{\boldmath $\tau$})=-iJ\int {\cal P}(h) dh\int d {\bf x} d {\bf y} \ e^{i{\bf y}.({\bf x}-{\bf x}^*)} {\bf y}.\mbox{\boldmath $\tau$}\prod_a \Theta(x^a \sigma^a)
\end{eqnarray}
and $F(\mbox{\boldmath $\sigma$})$ groups all other terms. The simple result follows:
\begin{eqnarray}
\label{Eq40a}
\Delta X_1=\sum_{\mbox{\boldmath $\sigma$}}e^{-\Lambda_0^*+gs+X_1(\mbox{\boldmath $\sigma$})}[I(\mbox{\boldmath $\sigma$},\mbox{\boldmath $1$})-I(\mbox{\boldmath $\sigma$},-\mbox{\boldmath $1$})] \nonumber\\
= -2iJ\sum_{\mbox{\boldmath $\sigma$}}e^{-\Lambda_0^* +gs+X_1(\mbox{\boldmath $\sigma$})}\int P(h) dh\int d {\bf x} \prod_a \Theta(x^a \sigma^a)\int d {\bf y} \ e^{i{\bf y}.({\bf x}-{\bf x}^*)} \sum_a y^a \nonumber\\ 
\ .
\end{eqnarray}
Using $-i\sum_a y^a=\partial e^{i{\bf y}.({\bf x}-{\bf x}^*)}/\partial x^*=\partial e^{i{\bf y}.({\bf x}-{\bf x}^*)}/\partial h$ and integrating by parts over $h$, we find that all $\sigma^a$'s must be equal in the sum over $\mbox{\boldmath $\sigma$}$, which yields
\begin{eqnarray}
\label{Eq43}
\Delta X_1=2J {\cal P}^*e^{-\Lambda_0^*}[e^{gn+X_1(\mbox{\boldmath $1$})}-e^{-gn+X_1(-\mbox{\boldmath $1$})}] 
\end{eqnarray}
where ${\cal P}^*$ is a shorthand for ${\cal P}(-H-J\Delta c_0^*)$. Eqs. (\ref{Eq35}), (\ref{Eq35a}), (\ref{Eq36a}), and (\ref{Eq43}) now form a closed system and  $\Lambda_0^*$, $\Delta c_0^*$, and $X_1(\pm\mbox{\boldmath $1$})$ can be calculated  in  the limit $n \rightarrow 0$ by expanding in powers of $n$ (i.e. assuming that $\Lambda_0^*=\Lambda_0^{*(0)}+n\Lambda_0^{*(1)}+...$, $\Delta c_0^*=\Delta c_0^{*(0)}+n\Delta c_0^{*(1)}+...$, and $X_1(\pm\mbox{\boldmath $1$)}$=$X_1^{(0)}(\pm\mbox{\boldmath
$1$)}$+$n X_1^{(1)}(\pm\mbox{\boldmath $1$)}...$).  At the lowest order in $n$, this gives
\begin{eqnarray}
\label{Eq44a}
\Lambda_0^{*(0)}&=0\\
X_1^{(0)}(\pm\mbox{\boldmath $1$)}&=0\\
\Delta c_0^{*(0)}&=2p(\Delta c_0^{*(0)})-1
\end{eqnarray}
where $p(m)=\int_{-H-Jm}^{\infty}{\cal P}(h)dh=(1/2)[1+\mbox{erf}([H+Jm]/\Delta \sqrt{2})]$. Note that, as stated before,  $\Lambda_0^{*(0)}=0$ is a prerequisite for a proper limit $n\rightarrow0$. In addition, Eq.  (\ref{Eq16}) readily gives 
\be
\label{Eq46}
m_0^*(H)=\lim_{n\rightarrow 0}\sum_{\mbox{\boldmath $\sigma$},\mbox{\boldmath $\tau$}}
\sigma^a c_0^*(\mbox{\boldmath $\sigma$},\mbox{\boldmath $\tau$})c_0^*(\mbox{\boldmath $\tau$},\mbox{\boldmath $\sigma$})=\Delta c_0^{*(0)} \ ,
\ee
so that $m_0^*$  satisfies the self-consistent mean-field equation\cite{S1993,DS1996,RTP2008}
\be
\label{Eq30a}
 m_0^*(H)=\mbox{erf}([H+Jm_0^*(H)]/\Delta \sqrt{2}) 
\ee 
and does not depend of $g$. 

At the order $n$, the solution of  Eqs. (\ref{Eq35}),(\ref{Eq35a}),(\ref{Eq36a}), and (\ref{Eq43}) yields $\Lambda_0^{*(1)}(g,H)=gm_0^*(H)$ (so that $m_0^*=\partial \Lambda_0^{*(1)}/\partial g$, as it must be), and the corresponding  complexity $\Sigma_{Q,0}(g,H)=\Lambda_0^{*(1)}(g,H)-m_0^*(H)g$ is thus identically zero. In the $H-m$ plane, this means that the quenched complexity is zero when $m= m_0^*(H)$, i.e. along the mean-field loop, and is not defined otherwise.  On the other hand, a  trivial calculation shows that the annealed complexity\cite{DRT2005} $\Sigma_A(m,H)=\lim_{N\to \infty}\frac{1}{N} \ln \overline{{\cal N}(m,H)}$ is also zero when 
$m= m_0^*(H)$ and strictly negative otherwise. We stress  that a zero complexity does not imply that there is a unique stable state in the thermodynamic limit.  The actual number of metastable states  along the curve $m_0^*(H)$ can be larger than $1$, even along the so-called ``unstable'' branch for $\Delta<\Delta_c^0$, as shown in Ref.\cite{RTP2008}.

The expressions of 
$\Delta c_0^{*(1)}$ and $X_1^{(1)}(\pm\mbox{\boldmath $1$)}$, which are needed in the subsequent calculations, are given at the end of Appendix A. 

\section{$1/z$ correction to the quenched complexity}

We now compute the $1/z$ correction to the complexity which, from Eq. (\ref{Eq17a}), is given by 
 \begin{equation}
\label{Eq64}
\Sigma_{Q,1}(g,H)= \Lambda_1^{*(1)}(g,H)-m_1^*(g,H)g \ .
\end{equation}
To  compute $\Lambda_1^{*(1)}(g,H)$ (and then $m_1^*(g,H)$ by derivation), we  have to fully solve  Eq. (\ref{Eq27}) so to obtain the explicit expressions of the  functions $I(\mbox{\boldmath $\sigma$},\mbox{\boldmath $\tau$})$ and $F(\mbox{\boldmath $\sigma$})$ defined by Eq. (\ref{Eq39}). After straightforward but lengthy algebraic manipulations (which are detailed in Appendix A), we find  that $\Lambda_1^*$ is given by
\begin{eqnarray}
\label{Eq61}
\Lambda_1^* =-\frac{1}{2}[X_1(\mbox{\boldmath $1$})^2c_0(\mbox{\boldmath $1$})+X_1(-\mbox{\boldmath $1$})^2c_0(-\mbox{\boldmath $1$})]-\frac{1}{2}[X_1(\mbox{\boldmath $1$})+X_1(-\mbox{\boldmath $1$})]-\frac{1}{8}
[X_1(\mbox{\boldmath $1$})+X_1(-\mbox{\boldmath $1$})]^2 \nonumber\\
-\frac{1}{2}J{\cal P}^{*}[X_1(\mbox{\boldmath $1$})-X_1(-\mbox{\boldmath $1$})][e^{-\Lambda_0^*+gn+X_1(\mbox{\boldmath $1$})}-e^{-\Lambda_0^*-gn+X_1(-\mbox{\boldmath $1$})}]\nonumber\\
-\frac{1}{2}J^2{\cal P}^{*2}[e^{-\Lambda_0^*+gn+X_1(\mbox{\boldmath $1$})}+e^{-\Lambda_0^*-gn+X_1(-\mbox{\boldmath $1$})}]^2 \nonumber\\
-J{\cal P}^*\Delta c_0^{*}\{[1+X_1(\mbox{\boldmath $1$})]e^{-\Lambda_0^*+gn+X_1(\mbox{\boldmath $1$})}-[1+X_1(-\mbox{\boldmath $1$})]e^{-\Lambda_0^*-gn+X_1(-\mbox{\boldmath $1$})}\}\nonumber\\
+\frac{J^2}{2}(1-\Delta c_0^{*2}){\cal P}^{'*}[e^{-\Lambda_0^*+gn+X_1(\mbox{\boldmath $1$})}-e^{-\Lambda_0^*-gn+X_1(-\mbox{\boldmath $1$})}]+\frac{1}{2}\sum_{\mbox{\boldmath $\sigma$}}[2J{\cal P}^*e^{-\Lambda_0^*+gs+X_1(\mbox{\boldmath $\sigma$})}]^2 \nonumber\\
\end{eqnarray}
and that $X_1(\mbox{\boldmath $\sigma$})$ satisfies the equation
\begin{eqnarray}
\label{Eq54}
X_1(\mbox{\boldmath $\sigma$})-X_1(\mbox{\boldmath $1$})=2J{\cal P}^*e^{-\Lambda_0^*}\big[e^{gs+X_1(\mbox{\boldmath $\sigma$})}-e^{gn+X_1(\mbox{\boldmath $1$})}] \ .
\end{eqnarray}
This equation plays a central role in our study and will be discussed in detail below. Let us first define the quantity $T(g,H)\equiv\sum_{\mbox{\boldmath $\sigma$}}[2J{\cal P}^*e^{-\Lambda_0^*+gs+X_1(\mbox{\boldmath $\sigma$})}]^2=\sum_{\mbox{\boldmath $\sigma$}}[X_1(\mbox{\boldmath $\sigma$})-X_1(\mbox{\boldmath $1$})+2J{\cal P}^*e^{-\Lambda_0^*+gn+X_1(\mbox{\boldmath $1$})}]^2$ and expand both sides of Eq. (\ref{Eq61}) in powers of $n$ (with $T(g,H)=T^{(0)}(g,H)+nT^{(1)}(g,H)+...$). Using Eqs. (40), (41), and (\ref{EqA15}), this yields
 \begin{equation}
\label{Eq62}
 \Lambda_1^{*(0)}(g,H)=-2J^2{\cal P}_0^{*2}+\frac{1}{2}T^{(0)}(g)
\end{equation}
and
\begin{eqnarray}
\label{Eq63}
\Lambda_1^{*(1)}(g,H)&=g(1-m_0^{*2})\frac{J^2{\cal P}_0^{'*}}{1-2J{\cal P}_0^*}(1-4\frac{J{\cal P}_0^*}{1-2J{\cal P}_0^*})+\frac{1}{2}T^{(1)}(g)
\end{eqnarray}
where ${\cal P}_0^*$ is a shorthand for  ${\cal P}(H+Jm_0^{*}(H))$ (i.e. ${\cal P}^*={\cal P}_0^*+0(n))$, and ${\cal P}'(h)=d{\cal P}(h)/dh$ (for brevity, the dependence of $T(g,H)$ on the field $H$ is dropped). From this we  obtain 
\begin{eqnarray}
\label{Eq65a}
m_1^*(g,H)&=\partial \Lambda_1^{*(1)}(g,H)/\partial g\nonumber\\
&=(1-m_0^{*2})\frac{J^2{\cal P}_0^{'*}}{1-2JP_0^*}(1-4\frac{J{\cal P}_0^*}{1-2J{\cal P}_0^*})+\frac{1}{2}\frac{\partial T^{(1)}(g)}{\partial g}
\end{eqnarray}
and
 \begin{equation}
\label{Eq65b}
\Sigma_{Q,1}(g,H)=\frac{1}{2}[T^{(1)}(g)-g\frac{\partial T^{(1)}(g)}{\partial g}] \ .
\end{equation}
(It can be checked through lengthy calculations that the same expression of $m_1^*(g,H)$ is recovered by expanding directly Eq. (\ref{Eq16})) to $O(1/z)$.)

The only remaining task is to solve Eq. (\ref{Eq54}) for $X_1(\mbox{\boldmath $\sigma$})$ and to calculate  $T^{(0)}(g)$ and $T^{(1)}(g)$. This equation can be conveniently rewritten as 
\begin{eqnarray}
\label{Eq65c}
W(\mbox{\boldmath $\sigma$})e^{W(\mbox{\boldmath $\sigma$})}= -A(g)e^{-A(g)+g(s-n)}
\end{eqnarray}
where $W(\mbox{\boldmath $\sigma$})= -X_1(\mbox{\boldmath $\sigma$})+X_1(\mbox{\boldmath $1$})-A(g)$ and $A(g)=2J{\cal P}^*e^{-\Lambda_0^*+gn+X_1(\mbox{\boldmath $1$})}$. 
Because of replica symmetry, $W(\mbox{\boldmath $\sigma$})$ only depends on $s=\sum_a\sigma^a$ and is  a function of the single variable $z(s,g)=-A(g)e^{-A(g)+g(s-n)}$. Eq. (\ref{Eq65c}) becomes
\begin{eqnarray}
\label{Eq66}
W(z)e^{W(z)}=z\ ,
\end{eqnarray}
which tells us that  $W(z)$ is the so-called ``Lambert function''\cite{CGHJK1996} (remarkably, this function already appears in the calculation of the average number of metastable states along the mean-field curve $m_0^*(H)$\cite{RTP2008}). The whole problem thus amounts to compute
\be
\label{Eq69}
T(g)=\sum_{\mbox{\boldmath $\sigma$}}W^2(z(s,g)) 
\ee
as $n \rightarrow 0$. 

Let us first consider  the  case $g=0$ which, as noted ealier, corresponds to the maximum of the complexity  and thus yields the total number of metastable states at the field $H$.  According to Eq. (\ref{Eq65a}), in order to obtain the corresponding magnetization $m^{max}(H)=m(g=0,H)$, one also needs to compute $\lim_{g\rightarrow 0}dT^{(1)}(g)/dg$. We then expand both sides of Eq. (\ref{Eq66}) in powers of $g$, using the formula for the $n$-th derivative of the Lambert function\cite{CGHJK1996}. This allows us to perform the sum over $\mbox{\boldmath $\sigma$}$  in each term of the expansion (computing $\sum_{\mbox{\boldmath $\sigma$}}s$, $\sum_{\mbox{\boldmath $\sigma$}}s^2$, etc...). The result is then expanded in powers of $n$, which gives
\begin{eqnarray}
\label{Eq70}
T^{(0)}(g)=t^2 \ ,
\end{eqnarray} 
and 
\begin{eqnarray}
\label{Eq71}
\fl T^{(1)}(g)=t^2\ln2+\frac{A^{(1)}(1-A^{(0)})}{A^{(0)}}\frac{2t^2}{1+t}g-\frac{2t^2}{1+t}g
+\frac{t^2(2+t)}{(1+t)^3}g^2
+\frac{t^2(2t^2+9t-8)}{6(1+t)^7}g^4+...\nonumber\\
&
\end{eqnarray} 
where $A(g)=A^{(0)}+nA^{(1)}g+...$ and $t=W(-A^{(0)}e^{-A^{(0)}})$.  Since $\Lambda_0^{*(0)}=X_1^{(0)}(\mbox{\boldmath $1$)}=0$ (Eqs. (\ref{Eq44a}) and (39)), one has 
\be
\label{Eq71a}
A^{(0)}=2J{\cal P}_0^* \ ,
\ee
and, using Eqs. (\ref{EqA15}),
\be
\label{Eq71b}
A^{(1)}=(1-m_0^*)\frac{2J{\cal P}_0^{*}}{1-2J{\cal P}_0^{*}}+2(1-m_0^{*2})\frac{J^2{\cal P}_0^{'*}}{[1-2J{\cal P}_0^*]^2}\ .
\ee

A question now needs to be addressed: which branch of $W(z)$ should be chosen? Recall that the Lambert function has two real branches, $W_0(z)$ and $W_{-1}(z)$, with a branch point at $z=-1/e$\cite{CGHJK1996}. The principal branch $W_0(z)$ takes on values between $-1$ to $+\infty$ for $z\ge-1/e$ (with $W_0(0)=0$) and $W_{-1}(z)$ takes on values between $-\infty$ and $-1$ for  $-1/e\le z\le 0$. Accordingly one has
\begin{eqnarray}
\label{Eq72}
W_0(-xe^{-x})&=-x \ \  \mbox{if $x\le 1$} \nonumber\\
W_0(-xe^{-x})&=-x' \ \  \mbox{if $x\ge 1$} 
\end{eqnarray} 
where $x'$ is the solution (smaller than 1) of the equation $x' e^{-x'}=xe^{-x}$. Similarly, 
\begin{eqnarray}
\label{Eq73}
W_{-1}(-xe^{-x})&=-x \ \  \mbox{if $x\ge 1$} \nonumber\\
W_{-1}(-xe^{-x})&=-x' \ \  \mbox{if $x\le 1$} 
\end{eqnarray} 
where $x'\ge 1$.

The choice of the branch is determined by  the fact that  $\Lambda^{*(0)}$ must vanish at all orders in $1/z$ in order to get a proper limit when $n\rightarrow 0$. Eq. (\ref{Eq62}) then implies that
\begin{equation}
\label{Eq74}
T^{(0)}(g)=4J^2{\cal P}_0^{*2}=A^{(0)2}
\end{equation}
whence $t=W(-A^{(0)}e^{-A^{(0)}})=-A^{(0)}$ in Eq. (\ref{Eq70}). Therefore, according to Eqs. (\ref{Eq72}) and (\ref{Eq73}), one has to choose the principal branch $W_0(z)$ for $A^{(0)}<1$ and the branch $W_{-1}(z)$ for $A^{(0)}>1$ (note that $W_0$ is the only branch that comes into play in the calculations of Ref.\cite{RTP2008}). The  series (\ref{Eq71}) then becomes
\begin{eqnarray}
\label{Eq75a}
T^{(1)}(g)=A^{(0)2} \ln 2+2A^{(0)}A^{(1)}g- \frac{2A^{(0)2}}{1-A^{(0)}}g+\frac{A^{(0)2}(2-A^{(0)})}{(1-A^{(0)})^3}g^2\nonumber\\
+\frac{A^{(0)2}(2A^{(0)2}-9A^{(0)}-8)}{6(1-A^{(0)})^7}g^4...
\end{eqnarray} 
so that, from Eqs. (\ref{Eq65a}) and (\ref{Eq65b}),
\begin{eqnarray}
\label{Eq75b}
m_1^{max}(H)=-m_0^*\frac{4J^2{\cal P}^{*2}_0}{1-2J{\cal P}^{*}_0}+
(1-m_0^{*2})\frac{J^2{\cal P}^{*'}_0}{1-2J{\cal P}^{*}_0} \ ,
\end{eqnarray}
and
\begin{equation}
\label{Eq75}
\Sigma_{Q,1}^{max}(H)\equiv\Sigma_{Q,1}(g=0,H)=2J^2{\cal P}_0^{*2}\ln 2 \ .
\end{equation}

This result will be commented on in the next section. More generally, the problem is that the series (\ref{Eq75a}) is strongly divergent when $g$ is large. We therefore need to find an integral representation that allows the sum to make sense for an arbitrary value of $g$. With this in mind, we write $z(s,g)=-A(g)e^{-A(g)}[1+(e^{g(s-n)}-1)]$ and  formally expand $W^2(z(s,g))$   around $-A(g)e^{-A(g)}$ in Eq. (\ref{Eq69}), which again allows us to perform the sum over $\mbox{\boldmath $\sigma$}$. Expanding the result in powers of $n$, we  get
\begin{eqnarray}
\label{Eq75d}
T^{(1)}(g)&=A^{(0)2}\ln2+g[\frac{A^{(1)}(1-A^{(0)})}{A^{(0)}}-1] z^{(0)}_0\frac{dW^2}{dz}\mid_{z=z^{(0)}_0}\nonumber\\
&+\sum_{p=1}^{\infty}\frac{z^{(0)p}_0}{p!}\frac{d^pW^2}{dz^p}\mid_{z=z^{(0)}_0}\sum_{q=1}^p (-1)^{p+q}\Big(_{q}^p\Big) \ln\cosh(qg) 
\end{eqnarray}
where $z^{(0)}_0=A^{(0)}e^{-A^{(0)}}$. The first few terms explicitly read
\begin{eqnarray} 
\label{Eq75e}
\fl T^{(1)}(g)=A^{(0)2}\ln 2+2A^{(0)}A^{(1)}g+\frac{2A^{(0)2}}{1-A^{(0)}}[\ln \cosh(g)-g]\nonumber\\
+\frac{A^{(0)2}(A^{(0)2}-A^{(0)}-1)}{(1-A^{(0)})^3}[2\ln \cosh(g)-\ln \cosh(2g)]\nonumber\\
+\frac{1}{3}\frac{A^{(0)3}(2A^{(0)3}-5A^{(0)2}+6)}{(1-A^{(0)})^5}[3\ln\cosh(g)-3\ln\cosh(2g)+\ln\cosh(3g)]... \nonumber\\ 
& 
\end{eqnarray}
This of course gives back the power series (\ref{Eq75a}) when expanding in powers of $g$.  To find an integral representation of (\ref{Eq75d}), we then use the identity\cite{PBM1986}
\begin{eqnarray}
\label{Eq76}
\tanh(qg) =\frac{2\pi}{g}\int_0^{\infty}dx \ \frac{\sin (2q\pi x)}{\sinh(\pi^2 x/g)}
\end{eqnarray}
which yields by integration
\begin{eqnarray}
\label{Eq76a}
\ln \cosh(qg)=\int_0^{\infty}dx \ \frac{1-\cos (2q\pi x)}{x\sinh(\pi^2 x/g)}
\end{eqnarray}
(the study is now restricted to $g>0$ but we know that in the end $\Sigma_{Q}(g,H)$ and $m^*(g,H)$ are even and odd functions of $g$, respectively).
As will be explained just below, it is convenient to split the domain of integration in Eq. (\ref{Eq76a}) into the successive intervals $[0,1]$, $[1,2]$, $[2,3]$, etc..., and  to change  $x$ to $1-x$ in the interval $[1/2,1]$. This yields
\begin{eqnarray}
\label{Eq77}
\ln \cosh(qg)=\int_0^{1/2} dx \ f(x,g)[1-\cos (2q\pi x)]
\end{eqnarray}
where
\begin{eqnarray}
\label{Eq78}
f(x,g)=\sum_{k=-\infty}^{\infty}\frac{1}{(x+k) \sinh(\pi^2(x+k)/g)} \ .
\end{eqnarray}
 Using $\sum_{q=1}^p(-1)^q\big(_{q}^p\big)=-1$, we then have
\begin{eqnarray}
\label{Eq79}
\sum_{q=1}^p (-1)^q\big(_{q}^p\big) \ln\cosh(qg)=-\int_0^{1/2} dx \ f(x,g) \Re(1-e^{-2i\pi x})^p \ ,
\end{eqnarray}
where $\Re$ denotes the real part, so that the series (\ref{Eq75d}) admits the following integral representation
\begin{eqnarray}
\label{Eq80}
T^{(1)}(g)&=A^{(0)2}\ln 2+2A^{(0)}A^{(1)}g-\frac{2A^{(0)2}}{1-A^{(0)}}g\nonumber\\
&+\int_0^{1/2} dx \ f(x,g)[A^{(0)2}-\Re \{W^2(-A^{(0)}e^{-A^{(0)}}e^{-2i\pi x})\}] \ , 
\end{eqnarray}
where we have interchanged summation and integration. (A priori  this is not justified since the series $\sum_{p=1}^{\infty}(-1)^p\frac{z^{(0)p}_0}{p!}\frac{d^pW^2}{dz^p}\mid_{z=z^{(0)}_0}(1-e^{-2i\pi x})^p$ is not uniformly convergent for $x$ in $[0,1/2]$; on the other hand, we have checked numerically that the power series (\ref{Eq75a}) is asymptotic to the result of Eq. (\ref{Eq80}) for sufficiently small $g$, even for $A^{(0)}>1$.) Note that it is essential that the path in the complex plane does not cross the branch cuts of $W_{0}$ and $W_{-1}$ when integrating over $x$ in Eq. (\ref{Eq80}). The branch cut of $W_{0}$ is $\{z:-\infty<z\le -1/e\}$ whereas $W_{-1}$ has the double branch-cut $\{z:-\infty<z\le -1/e\}$ and $\{z:-\infty<z\le 0\}$\cite{CGHJK1996}. Since $-A^{(0)}e^{-A^{(0)}}\ge -1/e$, the  restriction of  the domain of integration to the interval $[0,1/2]$ ensures that the imaginary part $\Im\{-A^{(0)}e^{-A^{(0)}}e^{-2i\pi x})\}\ge 0$, which explains that we use the integral representation (\ref{Eq77})  instead of (\ref{Eq76}).

Replacing $T^{(1)}(g)$ in Eq. (\ref{Eq65a}) and (\ref{Eq65b}) by its expression (70), and using Eqs. (\ref{Eq71a}) and (\ref{Eq71b}), we finally obtain
\begin{eqnarray}
\label{Eq81}
m_1^*(g,H)=-m_0^*\frac{4J^2{\cal P}^{*2}_0}{1-2J{\cal P}^{*}_0}+
(1-m_0^{*2})\frac{J^2{\cal P}^{*'}_0}{1-2J{\cal P}^{*}_0}+\frac{1}{2}\frac{\partial C(g,H)}{\partial g} \nonumber\\
\end{eqnarray}
and 
\begin{eqnarray}
\label{Eq81b}
\Sigma_{Q,1}(g,H)=2J^2{\cal P}^{*2}_0\ln 2+\frac{1}{2}(C(g,H)-g\frac{\partial C(g,H)}{\partial g})
\end{eqnarray}
where
\begin{eqnarray}
\label{Eq82}
C(g,H)&\equiv T^{(1)}(g)-A^{(0)2}\ln 2-2A^{(0)}A^{(1)}g+\frac{2A^{(0)2}}{1-A^{(0)}}g\nonumber\\
&= \int_0^{1/2} dx \ f(x,g)[A^{(0)2}-\Re \{W^2(-A^{(0)}e^{-A^{(0)}}e^{-2i\pi x})\}] \ .
\end{eqnarray}

Especially important are the limits $g\rightarrow \pm \infty$, as discussed in Ref.\cite{RTP2008}.  It turns out that further analytical progress can be made in this case.  Indeed, one can show that $f(x,g)\sim g/\sin^2(\pi x) -2\ln(2)$ when $g\rightarrow + \infty$. As a result, it is found that
\begin{eqnarray}
\label{Eq83}
m_1^*(g\rightarrow+ \infty,H)&=-m_0^*\frac{4J^2{\cal P}^{*2}_0}{1-2J{\cal P}^{*}_0}+
(1-m_0^{*2})\frac{J^2{\cal P}^{*'}_0}{1-2J{\cal P}^{*}_0}\nonumber\\
&+\frac{1}{2} \int_0^{1/2} dx \ \frac{A^{(0)2}-\Re \{W^2(-A^{(0)}e^{-A^{(0)}}e^{-2i\pi x})\}}{\sin^2(\pi x)} \ , 
\end{eqnarray}
and, after some additional manipulations, 
\begin{eqnarray}
\label{Eq83a}
\fl \Sigma_{Q,1}(g\rightarrow+ \infty,H)= \ln 2 \int_0^{1/2}dx \ \Re \{W^2(-A^{(0)}e^{-A^{(0)}}e^{-2i\pi x}) \}=-\frac{\ln 2}{2\pi} \Im\{\frac{w^3}{3}+\frac{w^2}{2}\}
\end{eqnarray}
where $w=W(+A^{(0)}e^{-A^{(0)}})$. 

With these expressions, we can now  study the behavior of the quenched complexity in the field-magnetization plane and compare with the hysteresis loop. (In Appendix C, we indicate a change of variable in Eqs. (\ref{Eq82}) and (\ref{Eq83}) which is better suited for numerical calculations.)

\section{Results and discussion}

The exact equations that describes the saturation hysteresis loop on a Bethe lattice were obtained in Ref.\cite{DSS1997} and the expansion of these equations to  $O(1/z)$ is worked out in Appendix B.  Let us recall again that the main feature is the existence of an out-of-equilibrium phase transition for $z\ge 4$ in the thermodynamic limit.  Whereas the hysteresis loop is smooth for $\Delta>\Delta_c(z)$, it has a jump for $\Delta<\Delta_c(z)$ that results from the flip of a finite fraction of spins (a macroscopic ``avalanche'').  The critical disorder $\Delta_c(z)$ increases  with $z$ (see e.g. Fig. 1 in Ref.\cite{ISV2006})  and the mean-field behavior described by Eq.(\ref{Eq30a}) is recovered when $z\rightarrow\infty$\cite{ISV2006}. From a mathematical point of view the transition for $\Delta<\Delta_c(z)$ is due to a saddle-node bifurcation\cite{DSS1997}: the self-consistent equation (\ref{EqB2}) (that corresponds to an increasing applied field) has three real roots in the range  $H_1(\Delta)<H<H_2(\Delta)$ and two of the solutions coalesce and become complex at the branching fields $ H_1$ and $H_2$  (see Fig. 3 below). As $H$ increases, the quantity  $U(H)$  defined by Eq. (\ref{EqB2}) follows initially the lowest solution and then  jumps to the highest one at $H=H_2(\Delta)$.  The symmetric behavior is observed when decreasing the  field. Whether or not the third ``unstable''  root and the corresponding reentrant branches of the hysteresis loop have a physical meaning will be discussed below. Note that  an intermediate branch is already present in the mean-field equation (\ref{Eq30a})  below the critical disorder $\Delta_c^0=J\sqrt{2/\pi}$\cite{DS1996}. Since $dm^*_0/dH<0$, this branch is also generally described as ``unstable'', which is actually a misleading terminology since some metastable states may be present, as shown in Ref.\cite{RTP2008}. The important point for the following discussion is that the  key quantity $A^{(0)}=2J{\cal P}(H+Jm_0^{*}(H))$ is smaller than $1$ for $\Delta>\Delta_c^0$ and on the stable portions of the mean-field curve for  $\Delta<\Delta_c^0$. On the other hand, it is larger  than $1$ along the intermediate ``unstable'' branch  ($A^{(0)}=1$ at the spinodal endpoints).

\subsection{$\Delta>\Delta_c^0$}

We first consider the strong-disorder regime in which the loop is smooth in the thermodynamic limit. For simplicity, we assume that $\Delta>\Delta_c^0>\Delta_c(z)$  so that the mean-field magnetization curve is also smooth. Then $A^{(0)}<1$ and $W=W_{0}$ in the equations of the preceding section. To illustrate numerically the general behavior, we take  $\Delta=1$  and $z=30$. 
\begin{figure}[ht]
\begin{center}
\epsfig{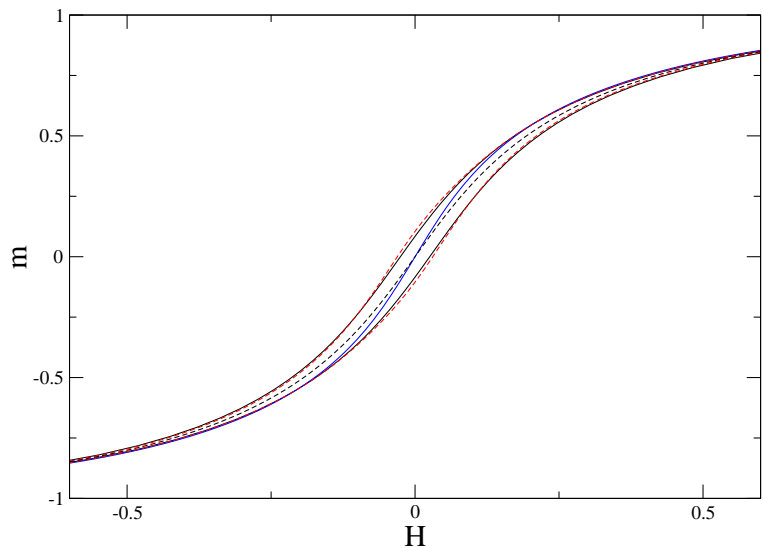}
\end{center}
\caption{\label{Fig1} Hysteresis loop for $\Delta=1$ and $z=30$ (solid black line).  At the scale of the figure, the $1/z$ expansion at first order (red dashed line) reproduces accurately the exact result. The solid blue line is the mean-field magnetization curve and the dashed black line is the locus of the maximum of the complexity.}
\end{figure}
\begin{figure}
\begin{center}
\epsfig{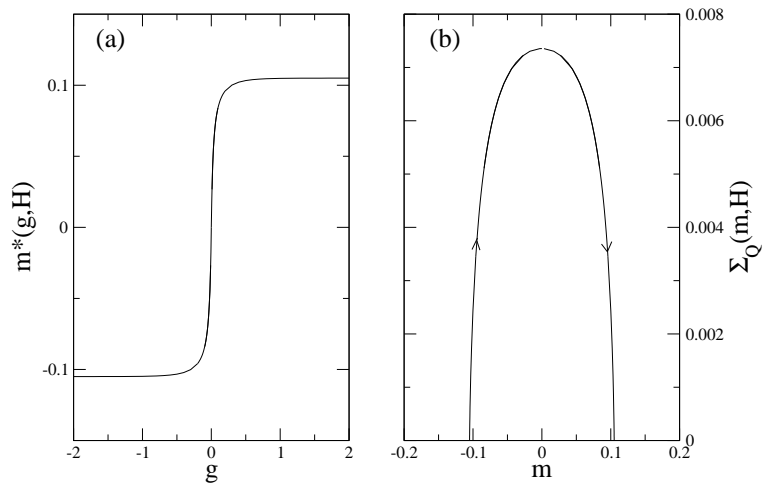}
\end{center}
\caption{\label{Fig2} Magnetization $m^*(g,H)=m_0^*(H)+(1/z)m_1^*(g,H)$ (left panel) and quenched complexity $\Sigma_Q(m,H)=(1/z)\Sigma_{Q,1}(m,H)$ (right panel) for 
$\Delta=1$, $z=30$, and $H=0$. The arrows in (b) indicate that $g$ increases from $-\infty$ and $+\infty$ along the curve. $\Sigma_Q(g,H)\rightarrow 0$ when $g\rightarrow\pm \infty$ so that  $\Sigma_Q(m,H)=0$ on the hysteresis loop (see text).}
\end{figure}

In Fig. 1, we first show the  hysteresis loop computed from the equations of Ref. \cite{DSS1997} (see also Eqs. (\ref{EqB1}) and (\ref{EqB2})) and the mean-field magnetization curve $m_0(H)$. We also plot the loop resulting from the expansion in $1/z$, $m(H)=m_0(H)+(1/z) m_1(H)$ where $m_1(H)$ is given by Eqs. (\ref{EqB16}) and (\ref{EqB17}). As can be seen,  the exact result is  accurately described by the first two terms in the $1/z$ expansion for the chosen value $z=30$.

The curves $m^*(g,H)=m_0^*(H)+(1/z)m_1^*(g,H)$ versus  $g$ and $\Sigma_Q(m,H)=(1/z)\Sigma_{Q,1}(m,H)$ versus $m$  following the results of  the preceding section are plotted in Figs. 2 (a) and 2 (b), respectively, for $H=0$. The results are qualitatively similar for any other value of the field. As $g$ varies from $-\infty$ to $+\infty$, the magnetization $m^*(g,H)$ increases monotonously, and the  complexity, as a function of $m$, has a shape similar to that observed for $z=2$\cite{DRT2005}. In particular, the slope is infinite when $g\rightarrow \pm \infty$, as a consequence of the Legendre relation $g=-\partial \Sigma_Q(m,H)/ \partial m$. 

By studying  the behavior as $g\rightarrow \pm \infty$, we can now answer the question that motivated the present  study: is the saturation hysteresis loop  the boundary of the region  in the  $H-m$ plane where the quenched complexity is positive ?
In the strong-disorder regime, the answer to this question is positive (at least at the order $1/z$). Indeed, when $W=W_0$,  the integral in the r.h.s. of Eq. (\ref{Eq83}) is  equal to $2A^{(0)2}/(1-A^{(0)})=8J^2{\cal P}_0^{*2}/(1-2J{\cal P}_0^{*})$. As a result, Eq. (\ref{Eq83})  identifies with Eq. (\ref{EqB17}) which describes the descending branch of the hysteresis loop. (Similarly, the ascending branch is recovered as $g \rightarrow -\infty$.) Moreover,  when $w=W_0(+A^{(0)}e^{-A^{(0)}})$, one has $\Im\{\frac{w^3}{3}+\frac{w^2}{2}\}=0$ in  Eq. (\ref{Eq83a}) so that $\Sigma_{Q,1}(g\rightarrow +\infty,H)$, the $1/z$ correction to the complexity along the hysteresis loop, is also zero.  These two results prove that the curve  $\Sigma_Q(m,H)=0$ coincides with the saturation hysteresis loop at this order. (These results can also be  obtained directly from  Eq. (\ref{Eq75d}) by expanding formally in powers of $A^{(0)}$, and taking the limit $g\rightarrow +\infty$: one  finds that  $T^{(1)}(g)\sim 2A^{(0)}A^{(1)}g +O(e^{-4g})$, which yields $C(g,H)\sim 2A^{(0)2}/(1-A^{(0)})g-A^{(0)2}\ln 2+O(e^{-4g})$.) Recall that a zero complexity only means that the number of metastable states is not exponentially growing with $N$, but it may still vary like a power law $N^{\alpha}$ with $N$.

\subsection{$\Delta<\Delta_c(z)$}

We now turn to the  low-disorder regime, illustrated by the case $\Delta=0.3$ and $z=100$. The hysteresis loop obtained from the equations of Ref.\cite{DSS1997} is  shown in Fig. 3 (for clarity, the mean-field magnetization curve $m_0(H)$ is only shown in Fig. 4 that zooms in the lower part of Fig. 3).
\begin{figure}[ht]
\begin{center}
\epsfig{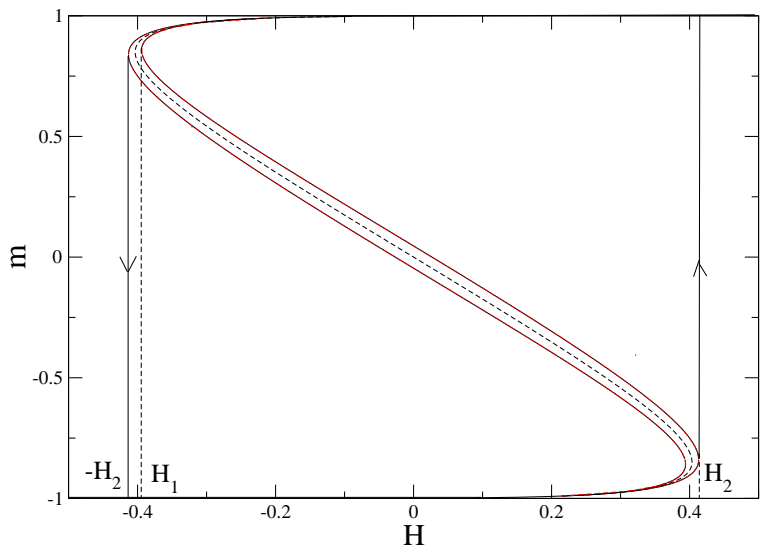}
\end{center}
\caption{\label{Fig3} Hysteresis loop for $\Delta=0.3$ and $z=100$ (solid black line)  showing jumps at the fields $H_2$ and $-H_2$. The two intermediate branches corresponding to the ``unstable'' root of the self-consistent equation of Ref. \cite{DSS1997}  are included.  The red dashed lines represent  the ``renormalized'' curves $H(m)=H_0(m)+(1/z)H_1(m)$ and the  dashed black line is the locus of the maxima of the complexity. $H_1$ and $H_2$ are the branching fields associated with Eq. (\ref{EqB2}) (see text). }
\end{figure}

The hysteresis loop has now a jump that occurs  at $H=H_2$ (resp. $-H_2$) when   increasing (resp. decreasing)  the applied field. For completeness we also plot in the figure the reentrant branches that correspond to the ``unstable'' root of the self-consistent equation of Ref. \cite{DSS1997}: for $H_1<H<H_2$ or $-H_2<H<-H_1$ when increasing or decreasing the field, respectively.  The red dashed lines represent the ``renormalized'' curves $H(m)=H_0(m)+H_1(m)/z+...$,  where $H_0(m)$ is the inverse function of $m_0(H)$ (i.e. $H_0(m_0(H))=H$) and $H_1(m)$ is given by Eq. (\ref{EqB18}). As explained at the end of Appendix B, this  ``renormalization''  of the $1/z$ expansion is necessary because the expansion $m(H)=m_0(H)+m_1(H)/z+...$ diverges when $H$ approaches the mean-field branching fields, that is when $A^{(0)}\rightarrow 1$. This problem can be circumvented  by working at constant magnetization instead of constant field and expanding $H$ instead of $m$. At the scale of Fig. 3, the resulting curves are indistinguishable from the exact ones,  including  the reentrant parts. We also display in the figure the locus of the maxima of the complexity which will be commented on later.
\begin{figure}
\begin{center}
\epsfig{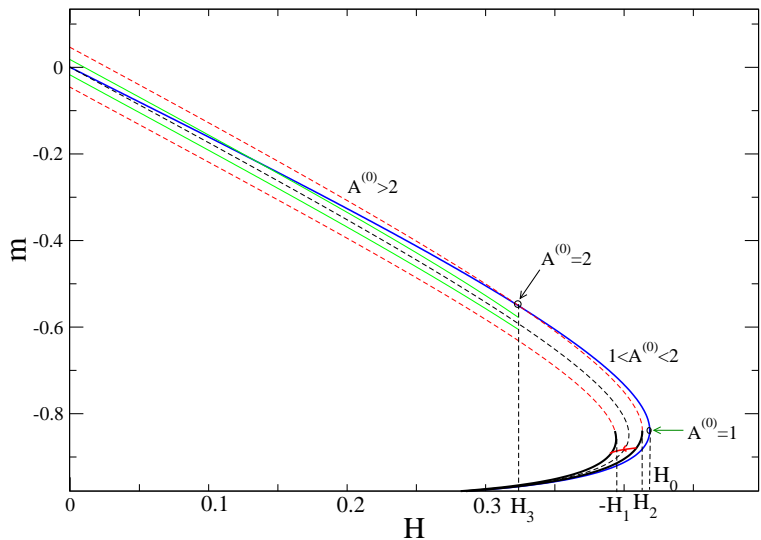}
\end{center}
\caption{\label{Fig4} Zoom on the lower part of Fig. 3. The blue curve is the mean-field magnetization and the arrows indicate the points on this curve  that correspond to $A^{(0)}=1$  ($H=H_0$) and $A^{(0)}=2$ ($H=H_3$). The complexity vanishes along the two ``stable'' branches in the lower part of the loop (bold solid lines) but remains finite along the two green curves in the central part of the loop for $0\le H\le H_3$. There is an exponential number of metastable states in the strip between the two green curves and this strip is narrower than the
region delimited by the dashed red lines. A first-order reversal curve is also shown in the lower part of the loop (solid red line).}
\end{figure}

We now discuss the results for the magnetization $m^*(g,H)$ and the  quenched complexity $\Sigma_Q(m,H)=(1/z)\Sigma_{Q,1}(m,H)$.  For clarity, we only focus on the lower part of the  loop (i.e. for $m<0$) as the upper one can be obtained via the symmetry $m(-H)=-m(H)$.  A priori, three values of the external field are noteworthy, as indicated in Fig. 4:  the two branching fields $-H_1\approx 0.391$ and $H_2\approx 0.413$, and the mean-field branching field $H_0\approx 0.418$ (note that $H_2$ is close to, but smaller than, $H_0$). As will be seen  below, the field  $H_3\approx 0.323$ such that $A^{(0)}(H_3)=2$, with $m_0^*(H_3)$ being the ``unstable'' solution of Eq. (\ref{Eq30a}),  will also play a  role.

We first consider the lower part of the loop for an applied  field $H<-H_1$.  ${\cal P}^{*}_0$ is then computed with the solution of  Eq. (\ref{Eq30a})  that corresponds to the lowest value of the magnetization $m_0^*$. Then $dm_0^*/dH>0$ and $A^{(0)}<1$  so that  $W=W_{0}$ in the equations of the preceding section. As a result, the qualitative behavior of $m^*(g,H)$ and $\Sigma_Q(m,H)$ is the same as in the strong-disorder regime: the  region between the two branches of the hysteresis contains an exponential number of metastable states and the quenched complexity  vanishes exactly along the two branches. This is illustrated  in Fig. 5 (a) and 5 (b) for $H=0.3$ ($m_0^*\approx -0.975$ and  $A^{(0)}\approx 0.210$). It must be emphasized that this part of the descending branch of the hysteresis (reached in the limit $g\rightarrow +\infty$) is physically accessible via a first-order reversal curve starting from the ascending branch, i.e. a protocol where $H$ is first increased to a value less than $H_2$ and then decreased. Such a curve is  shown in Fig. 4 (red line). States that can be reached in this way are called $H$-states\cite{BM2004} and it thus appears that there is an exponentially large number of  metastable states when $H$-states are present (most of the metastable states however are not $H$-states). 
\begin{figure}
\begin{center}
\epsfig{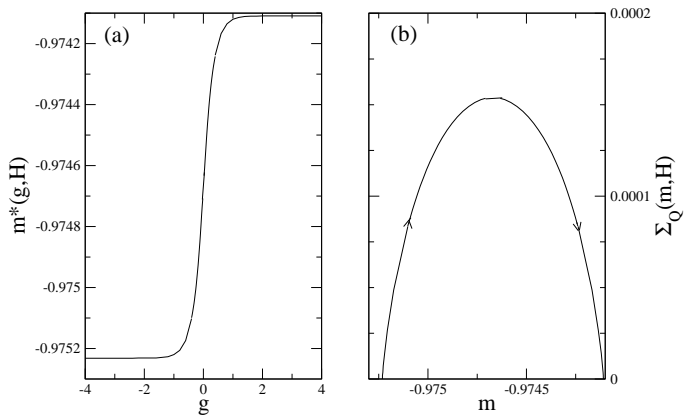}
\end{center}
\caption{\label{Fig5} Same as Fig. 2 for 
$\Delta=0.3$, $z=100$, and $H=0.3$ (lower part of the hysteresis loop).}
\end{figure}

The behavior is different in the central part of the loop, and we need to distinguish the two cases
$H<H_3$ and $H_3<H$. The first one is illustrated in Figs. 6 (a) and 6 (b) for  $H=0$.  ${\cal P}^{*}_0$ is then calculated with $m_0^*=0$ so that $A^{(0)}\approx 2.660>1$ and one has to take $W=W_{-1}$ in Eqs. (71)-(75). As can be seen in Fig. 6 (b), the main difference with the case $A^{(0)}<1$ is that the complexity does not vanish when $g\rightarrow \pm \infty$. (When setting $w=W_{-1}(+A^{(0)}e^{-A^{(0)}})$ in Eq. (\ref{Eq83a}),  $\Im\{\frac{w^3}{3}+\frac{w^2}{2}\}$ is no more equal to zero.) Moreover,  the integral in the r.h.s. of Eq. (\ref{Eq83}) is not equal to $2A^{(0)2}/(1-A^{(0)})$  so that Eq. (\ref{Eq83}) does not identify with  Eq. (\ref{EqB17}). As shown in Fig. 4, the actual region where the number of metastable states is exponentially large, bounded by the two curves  $m^*(H)=m_0^*(H)+(1/z) m_1^*(g\rightarrow \pm \infty,H)$ shown in green in the figure, is narrower than the region delimited by the reentrant branches of the hysteresis loop (red dashed lines in Fig. 4). This feature is in agreement with the numerical estimations of Ref.\cite{PRT2008} for $z=4$. Therefore, these reentrant branches, which correspond to the ``unstable''  root of the self-consistent equation of Ref.\cite{DSS1997}, bear no relation to the quenched complexity and do not seem to have a physical meaning. 

As $H$ increases from $0$ and $m_0^*(H)$  decreases along the  reentrant part of the mean-field  curve, $A^{(0)}$ decreases and the  minimum value of the complexity $\Sigma_Q(g\rightarrow \pm \infty,H)$ also decreases. Eventually, this  quantity  becomes negative for $A^{(0)}\approx 1.9885$, which corresponds to a field that is slightly larger than $H_3$. In fact, as soon as $H>H_3$, a new feature appears, which is illustrated in Figs. 7 (a) and 7 (b) for $H=0.36$ ($m_0^*=-0.627$ and $A^{(0)}\approx 1.787$): the magnetization $m^*(g,H)$ is no more a monotonously increasing  function of $g$; in other words, the mapping $\{m^{*}(g,H), \Sigma_Q(g,H)\} \mapsto  \Sigma_Q(m,H)$ is not unique. This induces the awkward behavior of $\Sigma_Q(m,H)$ observed in Fig. 7 (b).  Moreover, there is a whole range of $m$ where $\Sigma_Q(m,H)$ is negative. 
\begin{figure}
\begin{center}
\epsfig{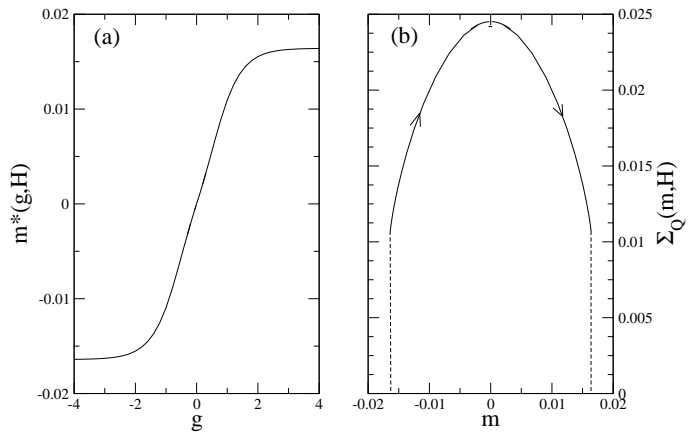}
\end{center}
\caption{\label{Fig6} Same as Fig. 2 for 
$\Delta=0.3$, $z=100$, and $H=0$ (central part of the hysteresis loop). Note that the complexity tends to a non-zero value ($ \approx 0.0107$) when $g\rightarrow\pm \infty$. There are no metastable states with $|m|\gtrsim0.016$.}
\end{figure}
\begin{figure}[ht]
\begin{center}
\epsfig{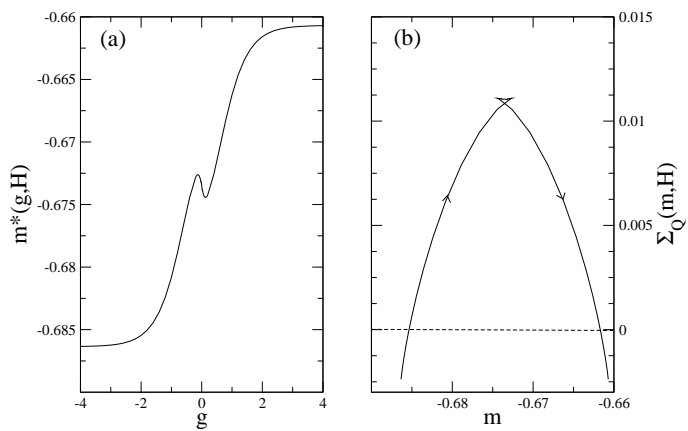}
\end{center}\caption{\label{Fig7} Same as Fig. 2 for $\Delta=0.3$, $z=100$, and $H=0.36$ (central part of the hysteresis loop).}
\end{figure}

The fact that $dm_1^*(g,H)/dg|_{g=0}=(1/2) \ d^2C(g,H)/dg^2|_{g=0}$  is negative for $1<A^{(0)}<2$ is already clear from Eq. (\ref{Eq75a}), the power series of $T^{(1}(g)$ (for $H=H_3$, $m^*(g,H)$ has an inflexion point at $g=0$). Unfortunately, this signals that the solution of the saddle-point equations is unstable or inconsistent and that there is a flaw in the analytical calculations of the preceding section. Indeed, $\Sigma_Q(m,H)$ must be always smaller than $\Sigma_{Q}^{max}(H)=\Sigma_Q(g=0,H)$, which implies that 
$\partial^2\Sigma_Q(g,H)/\partial g^2|_{m=cst.}=\partial^2\Lambda^{*(1)}(g,H)/\partial g^2=\partial m^*(g,H)/\partial g$ must be positive at $g=0$\cite{note2}.  It must be also emphasized that the quenched complexity is a quantity that cannot become negative. Despite our effort, we have not succeeded in finding the origin of the problem. We have considered various possibilities, including a breaking of symmetry of the replica vector $\bf g$ or the existence of complex saddles\cite{note3}, but unsuccessfully. Therefore, for $H_3<H<-H_1$, we  do not know the actual behavior of the complexity in the central region of the loop, and for $-H_1<H<H_2$, we can only predict that the complexity vanishes on the lower branch of the loop (since this branch is obtained with $A^{(0)}<1$ and $W=W_0$). The analytical behavior of the complexity in this intermediate region is obviously rather complicated.
\begin{figure}
\begin{center}
\epsfig{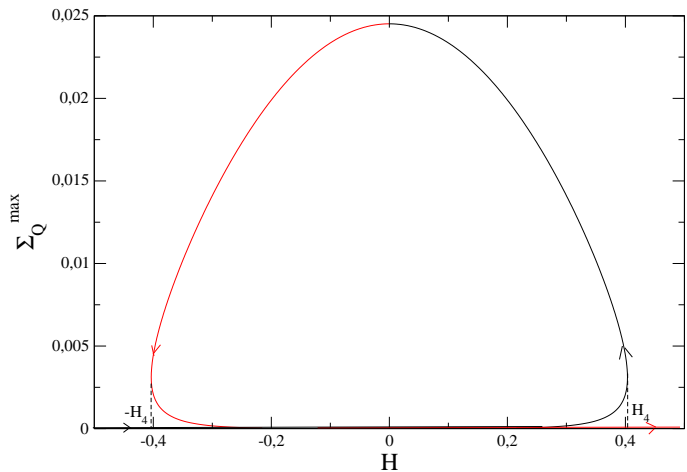}
\end{center}
\caption{\label{Fig8} Total complexity $\Sigma_{Q}^{max}(H)=(1/z)\Sigma_{Q,1}^{max}(H)$ for $\Delta=0.3$ and $z=100$ (see note \cite{note3}). As one follows the loop, the complexity increases from $0$ to $\approx 0.0245$ (black curve) and then decreases again to $0$ (red curve), as indicated by the arrows.}
\end{figure}

On the other hand, the value of the {\it total} quenched complexity $\Sigma_{Q}^{max}(H)=\Sigma_Q(g=0,H)$ is known in the whole $H-m$ plane (cf. Eq. (\ref{Eq75})), as well as the corresponding magnetization $m^{max}(H)$. These two quantities, properly ``renormalized''  in order to avoid a spurious divergence when $A^{(0)}\rightarrow 1$ \cite{note4}, are shown in Fig. 8 ($\Sigma_{Q}^{max}(H)$) and Figs. 3 and 4 ($m^{max}(H)$). $\Sigma_{Q}^{max}$ is multivalued for $-H_4<H<H_4$ where $H_4=(H_2-H_1)/2$ and it is maximum in the central part of the hysteresis loop. Therefore, the typical magnetization of the metastable states is actually a discontinuous function of $H$ with a negative slope around $H=0$. The surprising result that the magnetization of the most probable metastable states goes oppositely with the applied field is in full agreement with the numerical results of Ref.\cite{PRT2008} (see Figs. 16 and 17 in this reference).

Let us stress again that all the above results concern the behavior of the {\it quenched} complexity, associated with the typical number of metastable states. The annealed complexity is much more easily computed (along the lines of Ref.\cite{DRT2005}) but much less informative. It indeed remains positive outside the saturation hysteresis loop for finite $z$, as shown in Ref.\cite{DRT2005}, which signals the presence of ``atypical'' metastable states in this region. The two complexities only coincide  along the mean-field magnetization curve in the limit $z\rightarrow \infty$ and are  then equal to zero.

\section{Conclusion}

The main result of this paper concerns the organization of the metastable states of the RFIM in the field-magnetization plane and its relation with the saturation hysteresis loop. Calculations have been performed on a Bethe lattice of connectivity $z$ in the large-$z$ limit to order $1/z$. Although some pieces of the solution are still missing, the following conclusions can be made, that complete and confirm earlier computations:

1) When the hysteresis loop is smooth in the thermodynamic limit (strong-disorder regime), the quenched complexity $\Sigma_{Q}(m,H)$ is positive everywhere inside the  loop, i.e. the number of the  metastable states with a given magnetization increases exponentially with the system size, and it vanishes exactly along the loop. This is also the behavior computed analytically in one dimension\cite{DRT2005} and observed numerically for $z=4$ and on the cubic lattice\cite{PRT2008}. It is therefore quite reasonable to conclude that this is a general result. It would be nice, of course, to have another, more direct derivation of this property.  

2) When the hysteresis loop is discontinuous (low-disorder regime), the quenched complexity is positive in two sectors:

First, in  the two regions (before and after the jump in magnetization) that can be reached by a field history starting from one of the saturated states. It appears that there is an exponentially large number of  metastable states when $H$-states (i.e. field-reachable states\cite{BM2004}) are present, even if most of these metastables states are not $H$-states. The complexity vanishes along the two branches of the loop that coincide with the envelope of the $H$-states. In these two regions, the typical magnetization of the metastable states (i.e. the magnetization of the states whose number dominates the whole distribution) increases  with $H$. 

Second,  in a strip of finite width (of order $1/z$ in our calculations) around the mean-field magnetization curve. This  region in the middle of the loop is inaccessible to any field history  starting from the saturated states.  In addition, the complexity does not go continuously to zero on the borders of the strip, which moreover do not identify with the ``unstable'' branches computed from the fixed-point equations of Ref.\cite{DSS1997}.  It is in this strip that the number of metastable states is the largest and the typical magnetization  decreases with $H$.

\begin{figure}
\begin{center}
\epsfig{file=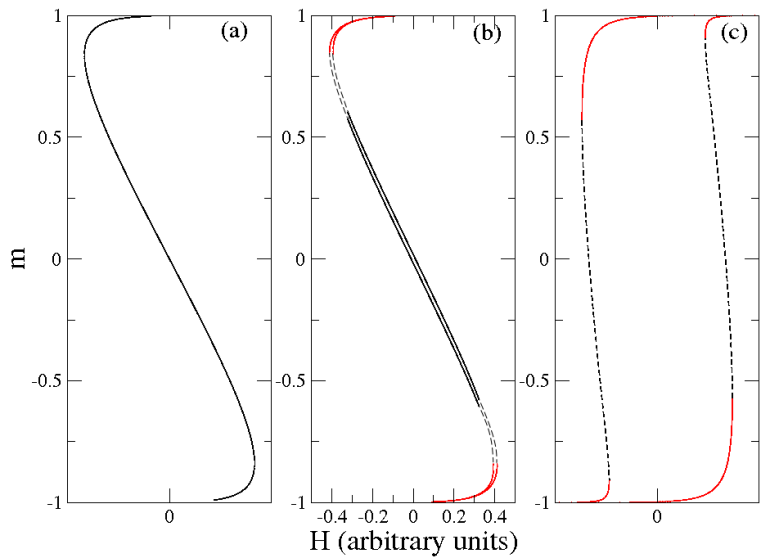, width=10cm,clip=}
\end{center}
\caption{\label{Fig9}  Evolution of  the envelope of the typical metastable states in the field-magnetization plane  in the low-disorder regime as a function of the connectivity $z$: (a) mean-field limit ($z=\infty$), (b) order $1/z$, (c) $z=4$. The complexity vanishes along the parts of the curves drawn in red. Certain parts are still putative (dashed lines). }
\end{figure}

The above results are  in agreement with the numerical study of Ref.\cite{PRT2008} for $z=4$ and for the cubic lattice and they suggest that in the low-disorder regime the region of positive (quenched) complexity evolves with connectivity as depicted in Fig. 9. We recall that a few metastable states are already present along the intermediate part of the mean-field curve\cite{RTP2008}. Although the precise behavior of the complexity in the vicinity of the ``knees'' is still unknown (and is certainly different on the Bethe and on euclidian lattices), it is now clear that the discontinuity in the  hysteresis loop associated with a macroscopic avalanche is due to the existence of a gap in the magnetization of the metastable states for a range of applied field. Consequently, by controlling the magnetization instead of the magnetic field\cite{IRV2006}, one should observe a reentrant loop, as indeed observed in some magnetic materials\cite{B1998} and other disordered systems\cite{BRIMPV2007,KPT2008}. Finally, one may wonder whether the most numerous, hence probable, states in the middle of the loop  are accessible dynamically; it is, however, dubious that this can be achieved by  performing a deep quench from $T=\infty$ to $T=0$ at a fixed field\cite{R2010}.

To conclude, we have  related the out-of-equilibrium disorder-induced transition in the RFIM at zero temperature to the distribution of the metastable states in the field-magnetization plane (thus replacing the dynamic problem by a purely static calculation). An interesting challenge would be to describe in a similar way other out-of-equilibrium field-driven transitions, for instance the depinning transition\cite{KR2000,PTZ2008}. In this case,  one must take explicitely into account the geometry of the lattice.

\ack

MLR  thanks T. Munakata for many fruitful discussions during his visit to Kyoto University.

\begin{appendix}

\section{Calculation of the complexity at the order $1/z$}

In this appendix, we detail the calculation of the complexity at the order $1/z$, solving Eq. (\ref{Eq27}).

We first note a difficulty in the $1/z$ expansion of the function $c^*(\mbox{\boldmath $\sigma$},\mbox{\boldmath $\tau$})$ arising from the presence of terms like ${\bf y}.\mbox{\boldmath $\tau$}$ in Eq. (\ref{Eq27}), $({\bf y}.\mbox{\boldmath $\tau$})^2$ at the next order, etc... These terms come from the expansion of $e^{-i\frac{J}{z}{\bf y}.\mbox{\boldmath $\tau$}}$ and they introduce successive derivatives of the Dirac $\delta$-function when integrating over $y^a$. This makes the  integrals ill-defined (or even divergent) and therefore this hierarchy of equations for $c_1^*(\mbox{\boldmath $\sigma$},\mbox{\boldmath $\tau$})$, $c_2^*(\mbox{\boldmath $\sigma$},\mbox{\boldmath $\tau$})$, etc... may just be considered as a formal and convenient way of book-keeping all the contributions at a certain order in $1/z$. To circumvent the difficulty, however, one can simply re-exponentiate the problematic terms. Consider for instance the quantity $I(\mbox{\boldmath $\sigma$},\mbox{\boldmath $\tau$})$ defined by Eqs. (\ref{Eq39}) and (\ref{Eq40}). It can be rewritten as
\begin{eqnarray}
\label{EqA1}
I(\mbox{\boldmath $\sigma$},\mbox{\boldmath $\tau$})&=\int dh {\cal P}(h) \int d {\bf x} d {\bf y}  e^{i{\bf y}.({\bf x}-{\bf x}^*)} \prod_a \Theta(x^a\sigma^a)\lim_{z\rightarrow \infty} z[e^{-i\frac{J}{z}{\bf y}.\mbox{\boldmath $\tau$}}-1]\nonumber\\
&=\lim_{z\rightarrow \infty}z\int dh {\cal P}(h) \{\prod_a \Theta(\sigma^a[x^*+\frac{J}{z}\tau^a])-\prod_a \Theta(\sigma^ax^*)\} \ ,
\end{eqnarray}
which readily yields
\begin{eqnarray}
\label{EqA2}
I(\pm\mbox{\boldmath $1$},\mbox{\boldmath $\tau$})&=J{\cal P}^*\big[(1\pm\tau)\prod_a \delta_K(\tau^a;\tau)-1\big]\nonumber\\
I(\mbox{\boldmath $\sigma$},\mbox{\boldmath $\tau$})&=2J{\cal P}^*\prod_a \delta_K(\sigma^a;\tau^a)
 \ \ \  \mbox{if \boldmath $\sigma$ $\neq \pm$ \boldmath $1$} \ .
\end{eqnarray}
where $\delta_K$ is the Kronecker symbol. This shows that the dependence of $c_1^*(\mbox{\boldmath $\sigma$},\mbox{\boldmath $\tau$})$ on $\mbox{\boldmath $\sigma$}$ and $\mbox{\boldmath $\tau$}$ is rather simple.

We now consider the quantity $X_1(\mbox{\boldmath $\sigma$})=\sum_{\mbox{\boldmath $\tau$}}c_1^*(\mbox{\boldmath $\tau$},\mbox{\boldmath $\sigma$})$. From Eq. (\ref{Eq39}), it satifies the self-consistent equation
\begin{eqnarray}
\label{EqA3}
X_1(\mbox{\boldmath $\sigma$})=\sum_{\mbox{\boldmath $\tau$}}F(\mbox{\boldmath $\tau$})+e^{-\Lambda_0^*} \sum_{\mbox{\boldmath $\tau$}}e^{gt+X_1(\mbox{\boldmath $\tau$})}I(\mbox{\boldmath $\tau$},\mbox{\boldmath $\sigma$}) 
\end{eqnarray} 
where $t=\sum_a \tau^a$. Using Eq. (\ref{EqA2}), we find after some algebra
\begin{eqnarray}
\label{EqA4}
X_1(\pm\mbox{\boldmath $1$})&=\sum_{\mbox{\boldmath $\tau$}}F(\mbox{\boldmath $\tau$})\pm J{\cal P}^*e^{-\Lambda_0^*}\big[e^{gn+X_1(\mbox{\boldmath $1$})}-e^{-gn+X_1(-\mbox{\boldmath $1$})}  \big]\nonumber\\
X_1(\mbox{\boldmath $\sigma$})&= \sum_{\mbox{\boldmath $\tau$}}F(\mbox{\boldmath $\tau$})+J{\cal P}^*e^{-\Lambda_0^*}\big[2e^{gs+X_1(\mbox{\boldmath $\sigma$})}-e^{gn+X_1(\mbox{\boldmath $1$})}-e^{-gn+X_1(-\mbox{\boldmath $1$})}\big]\ \ \  \mbox{if \boldmath $\sigma$ $\neq \pm$ \boldmath $1$} \nonumber\\
\end{eqnarray}
so that 
\begin{eqnarray}
\label{EqA5}
\sum_{\mbox{\boldmath $\tau$}}F(\mbox{\boldmath $\tau$})= \frac{1}{2}[X_1(\mbox{\boldmath $1$})+X_1(-\mbox{\boldmath $1$})]
\end{eqnarray}
and
\begin{eqnarray}
\label{EqA6}
X_1(\mbox{\boldmath $\sigma$})= X_1(\mbox{\boldmath $1$})+2J{\cal P}^*e^{-\Lambda_0^*}\big[e^{gs+X_1(\mbox{\boldmath $\sigma$})}-e^{gn+X_1(\mbox{\boldmath $1$})}] \ .
\end{eqnarray}
This is the crucial self-consistency equation (\ref{Eq54}), valid for all $\mbox{\boldmath $\sigma$}$.

We then calculate the quantity $F(\mbox{\boldmath $\sigma$})$ defined by Eq.(\ref{Eq39}). After some manipulations, we obtain from Eqs. (21b), (\ref{Eq27}) and (\ref{Eq39})
\begin{eqnarray}
\label{EqA7}
F(\mbox{\boldmath $\sigma$})&=[-\Lambda_1^*-X_1(\mbox{\boldmath $\sigma$})+X_2(\mbox{\boldmath $\sigma$})-\frac{1}{2}X_1(\mbox{\boldmath $\sigma$})^2]c_0^*(\mbox{\boldmath $\sigma$})\prod_{a}\delta_K(\sigma^a;\sigma)\nonumber\\
&+e^{-\Lambda_0^* +gs+X_1(\mbox{\boldmath $\sigma$})}\sum_{\mbox{\boldmath $\tau$}}I(\mbox{\boldmath $\sigma$},\mbox{\boldmath $\tau$})c_1^*(\mbox{\boldmath $\tau$},\mbox{\boldmath $\sigma$})\nonumber\\
&+e^{-\Lambda_0^* +gs+X_1(\mbox{\boldmath $\sigma$})} \int {\cal P}(h) dh\int d {\bf x} d {\bf y} \ e^{i{\bf y}.({\bf x}-{\bf x}^*)} \prod_a \Theta(x^a \sigma^a)\Big\{iJ\Delta c_0^{*}[1+X_1(\mbox{\boldmath $\sigma$})]\nonumber\\
&\times\sum_a y^a-\frac{J^2}{2}(1-\Delta c_0^{*2})(\sum_a y^a)^2\Big\} 
\end{eqnarray}
where  $X_2(\mbox{\boldmath $\sigma$})=\sum_{\mbox{\boldmath $\tau$}}c_2^*(\mbox{\boldmath $\tau$},\mbox{\boldmath $\sigma$})$. The last term is computed by integration by parts over $h$, which gives
\begin{eqnarray}
\label{EqA8}
F(\mbox{\boldmath $\sigma$})&=[-\Lambda_1^*-X_1(\mbox{\boldmath $\sigma$})+X_2(\mbox{\boldmath $\sigma$})-\frac{1}{2}X_1(\mbox{\boldmath $\sigma$})^2]c_0^*(\mbox{\boldmath $\sigma$}) \prod_{a}\delta_K(\sigma^a;\sigma)\nonumber\\
&+e^{-\Lambda_0^* +gs+X_1(\mbox{\boldmath $\sigma$})}\sum_{\mbox{\boldmath $\tau$}}I(\mbox{\boldmath $\sigma$},\mbox{\boldmath $\tau$})c_1^*(\mbox{\boldmath $\tau$},\mbox{\boldmath $\sigma$})\nonumber\\ 
&+\sigma e^{-\Lambda_0^* +gs+X_1(\mbox{\boldmath $\sigma$})}\{-J\Delta c_0^{*}[1+X_1(\mbox{\boldmath $\sigma$})]{\cal P}^*
+\frac{J^2}{2}(1-\Delta c_0^{*2}){\cal P}^{'*}\}\prod_{a}\delta_K(\sigma^a;\sigma)\nonumber\\
\end{eqnarray}
where ${\cal P}'(h)$ is the derivative of ${\cal P}(h)$ with respect to $h$.
It thus remains to calculate $\sum_{\mbox{\boldmath $\tau$}}I(\mbox{\boldmath $\sigma$},\mbox{\boldmath $\tau$})c_1^*(\mbox{\boldmath $\tau$},\mbox{\boldmath $\sigma$})$ and $X_2(\mbox{$\pm$\boldmath $1$})=\sum_{\mbox{\boldmath $\tau$}}c_2^*(\mbox{\boldmath $\tau$},\mbox{$\pm$\boldmath $1$})$. Using Eqs. (\ref{Eq39}) and (\ref{EqA2}), we find after some algebra
\begin{eqnarray}
\label{EqA9}
\sum_{\mbox{\boldmath $\tau$}}I(\mbox{$\pm$ \boldmath $1$},\mbox{\boldmath $\tau$})c_1^*(\mbox{\boldmath $\tau$},\mbox{$\pm$\boldmath $1$})&=J{\cal P}^*[2F(\pm \mbox{\boldmath $1$})-X_1(\pm\mbox{\boldmath $1$})+2J{\cal P}^{*}e^{-\Lambda_0^*\pm gn+X_1(\pm\mbox{\boldmath $1$})}]\nonumber\\
\sum_{\mbox{\boldmath $\tau$}}I(\mbox{\boldmath $\sigma$},\mbox{\boldmath $\tau$})c_1^*(\mbox{\boldmath $\tau$},\mbox{\boldmath $\sigma$})&=2J{\cal P}^*F( \mbox{\boldmath $\sigma$})+4J^2P^{*2}e^{-\Lambda_0^*+gs+X_1(\mbox{\boldmath $\sigma$})}\nonumber\\
&=2J{\cal P}^*[F( \mbox{\boldmath $\sigma$})-W(\mbox{\boldmath $\sigma$})]\ \ \  \mbox{if \boldmath $\sigma$ $\neq \pm$ \boldmath $1$} 
\end{eqnarray}
where $W(\mbox{\boldmath $\sigma$})=-X_1(\mbox{\boldmath $\sigma$})+X_1(\mbox{\boldmath $1$})-A(g)$ and $A(g)=2J{\cal P}^*e^{-\Lambda_0^*+gn+X_1(\mbox{\boldmath $1$})}$. Inserting  into Eq. (\ref{EqA7}), we obtain
\begin{eqnarray}
\label{EqA10}
\fl [1-2J{\cal P}^*e^{-\Lambda_0^* \pm gn+X_1(\pm \mbox{\boldmath $1$})}]F( \mbox{$\pm$ \boldmath $1$})=[-\Lambda_1^*-X_1(\mbox{$\pm$\boldmath $1$})+X_2(\mbox{$\pm$\boldmath $1$})-\frac{1}{2}X_1(\mbox{$\pm$\boldmath $1$})^2]c_0^*(\mbox{$\pm$\boldmath $1$})\nonumber\\
-J{\cal P}^*e^{-\Lambda_0^* \pm gn+X_1(\mbox{$\pm$\boldmath $1$})}[X_1(\pm\mbox{\boldmath $1$})-2J{\cal P}^{*}e^{-\Lambda_0^*\pm gn+X_1(\pm\mbox{\boldmath $1$})}]\nonumber\\ 
\pm e^{-\Lambda_0^* \pm gn+X_1(\mbox{$\pm$\boldmath $1$})}\{-J\Delta c_0^{*}[1+X_1(\mbox{$\pm$\boldmath $1$})]{\cal P}^*
+\frac{J^2}{2}(1-\Delta c_0^{*2}){\cal P}^{*'}\}
\end{eqnarray}
and
\begin{eqnarray}
\label{EqA11}
\fl [1-2J{\cal P}^*e^{-\Lambda_0^*+gs+X_1(\mbox{\boldmath $\sigma$})}]F( \mbox{\boldmath $\sigma$})&=[2J{\cal P}^*e^{-\Lambda_0^*+gs+X_1(\mbox{\boldmath $\sigma$})}]^2 
=W^2(\mbox{\boldmath $\sigma$})\ \ \  \mbox{if \boldmath $\sigma$ $\neq \pm$ \boldmath $1$} 
\end{eqnarray} 
where we have used Eq. (\ref{EqA6}) and the definition of $W(\mbox{\boldmath $\sigma$})$.

In order to compute $X_2(\mbox{$\pm$\boldmath $1$})$, we consider the normalization equation (\ref{Eq15}) at the order $1/z^2$,  which reads
\be
\label{EqA12}
2\sum_{\mbox{\boldmath $\sigma$},\mbox{\boldmath $\tau$}}c_0^*(\mbox{\boldmath $\sigma$},\mbox{\boldmath $\tau$})c_2^*(\mbox{\boldmath $\tau$},\mbox{\boldmath $\sigma$})+\sum_{\mbox{\boldmath $\sigma$},\mbox{\boldmath $\tau$}}c_1^*(\mbox{\boldmath $\sigma$},\mbox{\boldmath $\tau$})c_1^*(\mbox{\boldmath $\tau$},\mbox{\boldmath $\sigma$})=0 \ ,
\ee
whence
\begin{eqnarray}
\label{EqA13}
c_0(\mbox{\boldmath $1$})X_2(\mbox{\boldmath $1$})+c_0(-\mbox{\boldmath $1$})X_2(-\mbox{\boldmath $1$})=-\frac{1}{2}\sum_{\mbox{\boldmath $\sigma$},\mbox{\boldmath $\tau$}}c_1^*(\mbox{\boldmath $\sigma$},\mbox{\boldmath $\tau$})c_1^*(\mbox{\boldmath $\tau$},\mbox{\boldmath $\sigma$}) \ .
\end{eqnarray}
Using Eqs. (\ref{Eq39}), (\ref{EqA5}) and (\ref{EqA11}), we find 
\begin{eqnarray}
\label{EqA14}
\fl c_0(\mbox{\boldmath $1$})X_2(\mbox{\boldmath $1$})+c_0(-\mbox{\boldmath $1$})X_2(-\mbox{\boldmath $1$})=
F(\mbox{\boldmath $1$})[1-2J{\cal P}^*e^{-\Lambda_0^*+gn+X_1(\mbox{\boldmath $1$})}]+F(-\mbox{\boldmath $1$})[1-2J{\cal P}^*e^{-\Lambda_0^*-gn+X_1(-\mbox{\boldmath $1$})}]\nonumber\\
-\frac{1}{8}[X_1(\mbox{\boldmath $1$})+X_1(-\mbox{\boldmath $1$})]^2
-\frac{1}{2}[X_1(\mbox{\boldmath $1$})+X_1(-\mbox{\boldmath $1$})][1-J{\cal P}^*e^{-\Lambda_0^*}(e^{gn+X_1(\mbox{\boldmath $1$})}+e^{-gn+X_1(-\mbox{\boldmath $1$})})]\nonumber\\
-J^2{\cal P}^{*2}\big[\frac{5}{2}e^{2[-\Lambda_0^*+gn+X_1(\mbox{\boldmath $1$})]}+\frac{5}{2}e^{2[-\Lambda_0^*-gn+X_1(-\mbox{\boldmath $1$})]}+e^{-2\Lambda_0^*+X_1(\mbox{\boldmath $1$})+X_1(-\mbox{\boldmath $1$})}\big]\nonumber\\
+\frac{1}{2}T(g)
\end{eqnarray}
where we have introduced $T(g)=\sum_{\mbox{\boldmath $\sigma$}}W^2(\mbox{\boldmath $\sigma$})$ (cf. Eq. (\ref{Eq69})).

Inserting this into Eqs.(\ref{EqA10}), adding the two equations for $F(\mbox{\boldmath $1$})$ and $F( \mbox{-\boldmath $1$})$, and using the normalization equation (\ref{Eq30}),  we finally obtain
Eq. (\ref{Eq61}).
(Note that, remarkably, the terms involving $F( \mbox{$\pm$ \boldmath $1$})$ have cancelled out so that there is no need to compute $X_2(\mbox{\boldmath $1$})$ and $X_2(-\mbox{\boldmath $1$})$ separately, which would imply to also consider the equation for $c_2(\mbox{\boldmath $\sigma$},\mbox{\boldmath $\tau$})$.) 

In order to calculate $\Lambda_1^{*(1)}$, one also needs the following expressions of $\Delta c_0^{*(1)}$ and $X_1^{(1)}(\pm\mbox{\boldmath $1$)}$, solutions of Eqs. (\ref{Eq35}), (\ref{Eq35a}), (\ref{Eq36a}), and (\ref{Eq43}) at the order $n$,

\begin{eqnarray}
\label{EqA15}
\Delta c_0^{*(1)}&=g\frac{1-m_0^{*2}}{[1-2J{\cal P}_0^*]^2} \nonumber\\
X_1^{(1)}(\pm\mbox{\boldmath $1$)}&=\pm g (1\pm m_0^{*})\frac{2J{\cal P}_0^*}{1-2J{\cal P}_0^*}\ .
\end{eqnarray}

\section{Calculation of the hysteresis loop at the order $1/z$}

In this Appendix, we compute the hysteresis loop at the order $1/z$, starting from the equations derived 
by Dhar et al.\cite{DSS1997}. The approach to the mean-field behavior has been studied numerically in Ref.\cite{ISV2006}, but, as far as we know, the explicit calculation has only been performed in the limit $z\rightarrow \infty$. 
According to Ref.\cite{DSS1997}, the  magnetization $m(H)$ along the ascending branch of the hysteresis loop is given by 
\be
\label{EqB1}
\frac{1}{2}[m(H)+1]=\sum_{k=0}^{z}\big(_k^z\big) U^{k}(1-U)^{z-k}p_k  
\ee
where $U(H)$ is solution of the self-consistent equation
\be
\label{EqB2}
U=\sum_{k=0}^{z-1}\big(_k^{z-1}\big) U^{k}(1-U)^{z-1-k}p_k
\ee
and $p_k(H)=\int_{-H+J(1-\frac{2k}{z})}^{+\infty}{\cal P}(h)dh$ is the probability for a spin down to flip up at the field $H$ when $k$ of its nearest neighbours are up\cite{DSS1997} ($J$ has been rescaled by $z$). In order to compute the $O(1/z)$ correction, we assume the expansion  $U= U_0+U_1/z+U_2/z^2+...$, and, anticipating the fact that the sums in Eqs. (\ref{EqB1}) and (\ref{EqB2}) are dominated by the term corresponding to $k=zU_0$, we change to the variable $t=k/z$ and replace the sums by integrals that can be computed asymptotically by the Laplace method\cite{BO1987}. Eq. (\ref{EqB1}) then becomes
\begin{eqnarray}
\label{EqB3}
\fl \frac{1}{2}(m(H)+1)\simeq z\int_0^1 dt \exp[\ln \big(_{zt}^z\big) +zt\ln U+z(1-t)\ln(1-U)]p(t)
\end{eqnarray}
with $p(t)=\int_{-H+J(1-2t)}^{\infty}{\cal P}(h)dh$.  Similarly, Eq. (\ref{EqB2}), which is conveniently  rewritten as $U(1-U)=\sum_{k=0}^{z}\frac{z-k}{z}\big(_k^{z}\big)U^{k}(1-U)^{z-k}p_k$, becomes 
\begin{eqnarray}
\label{EqB4}
\fl U(1-U)\simeq z\int_0^1 dt (1-t)\exp[\ln \big(_{zt}^{z}\big)+zt\ln U+z(1-t)\ln(1-U)]p(t) \ .
\end{eqnarray}
Expanding $U$ and $m(H)$, and using the Stirling approximation for the binomial coefficient in the large-z limit, we find 
\begin{eqnarray}
\label{EqB5}
\fl \frac{1}{2}(m_0(H)+1)+\frac{1}{2z}m_1(H)+......&\simeq \sqrt{\frac{z}{2\pi }}\int_0^1 dte^{z\phi (t)} f(t)[1+\frac{1}{z}g(t)+...]
\end{eqnarray}
and
\begin{eqnarray}
\label{EqB6}
 \fl U_0(1-U_0)+\frac{1}{z}U_1(1-2U_0)+...\simeq \sqrt{\frac{z}{2\pi }}\int_0^1 dte^{z\phi (t)} (1-t)f(t)[1+\frac{1}{z}g(t)+...]
\end{eqnarray}
where
\begin{eqnarray}
\label{EqB7}
\fl \phi (t)=t\ln \frac {U_0}{t}+(1-t)\ln \frac {1-U_0}{1-t}
\end{eqnarray}
\begin{eqnarray}
\label{EqB8}
\fl f(t)=\frac{p(t)}{\sqrt{t(1-t)}}\exp[t\frac{U_1}{U_0}-(1-t)\frac{U_1}{1-U_0}]
\end{eqnarray}
\begin{eqnarray}
\label{EqB9}
\fl g(t)=t[\frac{U_2}{U_0}-\frac{U_1^{2}}{2U_0^{2}}]-(1-t)[\frac{U_2}{1-U_0}+\frac{U_1^{2}}{2(1-U_0)^2}]+\frac{1}{12}[1-\frac{1}{t(1-t)}] \ .
\end{eqnarray}
The function $\phi(t)$ is maximum for $t=U_0$, with $\phi(U_0 )=0$, $[-\phi''(U_0)]^{-1/2}=\sqrt{U_0(1-U_0)}$ and $f(U_0)=p(U_0)/\sqrt{U_0(1-U_0)}$, so that, at the leading order, one obtains from Eq. (\ref{EqB5})
\begin{eqnarray}
\label{EqB10}
\frac{1}{2}(m_0(H)+1)=p(U_0)
\end{eqnarray}
and, from Eq. (\ref{EqB6}), 
\begin{eqnarray}
\label{EqB11}
U_0=p(U_0)
\end{eqnarray}
Using the definition of $p(t)$, this  readily yields Eq. (\ref{Eq30a}), the self-consistent mean-field equation.
To compute the terms of order $1/z$ we need to take into account the first correction to the asymptotic behavior in the Laplace'method\cite{BO1987}. This gives from Eq. (\ref{EqB5})
\begin{eqnarray}
\label{EqB12}
\fl \frac{1}{2}m_1(H)= \sqrt{U_0(1-U_0)}\ \Big\{f(U_0)g(U_0) -\frac{f''(U_0)}{2\phi''(U_0 )}+ \frac{f(U_0)\phi''''(U_0 )}{8[\phi''(U_0)]^2}+\frac{f'(U_0)\phi'''(U_0 )}{2[\phi''(U_0)]^2}\nonumber\\
-\frac{5f(U_0)[\phi'''(U_0 )]^2}{24[\phi''(U_0)]^3}\Big\}\nonumber\\
\end{eqnarray}
and, from Eq. (\ref{EqB6}), a similar expression for $U_1(1-2U_0)$ with $f(t)$ replaced by $(1-t)f(t)$. Using $U_0=(m_0+1)/2$, we have
\label{EqB13}
\begin{eqnarray}
-\frac{1}{2\phi''( U_0)}=\frac{1-m_0^2}{8}
\end{eqnarray}
\begin{eqnarray}
\frac{\phi''''(U_0 )}{8[\phi''(U_0)]^2}=-\frac{1+3m_0^2}{4(1-m_0^2)}
\end{eqnarray}
\begin{eqnarray}
\frac{\phi'''( U_0)}{2[\phi''(U_0)]^2}=-\frac{m_0}{2}
\end{eqnarray}
\begin{eqnarray}
-\frac{5[\phi'''(U_0 )]^2}{24[\phi''(U_0)]^3}=\frac{5m_0^2}{6(1-m_0^2)} \ .
\end{eqnarray}
After some algebra, we obtain
\begin{eqnarray}
\label{EqB14}
m_1=4J{\cal P}_0^*U_1+(1-m_0^2)J^2{\cal P}_0^{*'}
\end{eqnarray}
and
\begin{eqnarray}
\label{EqB15}
U_1=-(1+m_0)\frac{J{\cal P}_0^{*}}{1-2J{\cal P}_0^{*}}+\frac{1-m_0^2}{2}J{\cal P}_0^{*'} \ ,
\end{eqnarray}
where $2J{\cal P}_0^*\equiv 2J {\cal P}(H+Jm_0)=p'(c)$ and ${\cal P}_0^{*'}=P'(H+Jm_0)$, using the notations of sections III 	and IV (with $m_0^*$ replaced by $m_0$).  Hence finally,

\begin{eqnarray}
\label{EqB16}
m_1(H)=-(1+m_0)\frac{4J^2{\cal P}_0^{*2}}{1-2J{\cal P}_0^{*}}+(1-m_0^2)\frac{J^2{\cal P}_0^{*'}}{1-2J{\cal P}_0^{*}} \ .
\end{eqnarray}

Similarly, along the descending branch (using the symmetry $H \rightarrow -H$, $m \rightarrow -m$), 
\begin{eqnarray}
\label{EqB17}
m_1(H)=(1-m_0)\frac{4J^2{\cal P}_0^{*2}}{1-2J{\cal P}_0^{*}}+(1-m_0^2)\frac{J^2{\cal P}_0^{*'}}{1-2J{\cal P}_0^{*}} \ .
\end{eqnarray}

For $\Delta<\Delta_c^0=J \sqrt{2/\pi}$, the function $m_0(H)$ is multivalued and Eqs. (\ref{EqB16}) and (\ref{EqB17}) diverge at the mean-field spinodal where $\partial m_0/\partial H=2{\cal P}_0^*/ ({1-2J{\cal P}_0^{*}})\rightarrow \infty$. This problem may be cured by considering 
the field as a function of the magnetization and expanding $H$ as $H(m)=H_0(m)+H_1(m)/z+H_2(m)/z^2+...$, where $H_0(m)$ is the inverse function of $m_0(H)$, the solution of Eq. (\ref{Eq30a}).  Then $H_1(m)=-m_1(H_0(m))/[\partial m_0/\partial H\vert_{H=H_0(m)}]=-m_1(H_0(m))[1-2J{\cal P}_0^*(H_0(m)]/2{\cal P}_0^*(H_0(m))$ and from Eq. (\ref{EqB16}) we obtain 
\begin{eqnarray}
\label{EqB18}
H_1(m)=2(1+m)J^2{\cal P}_0^*(m)-\frac{1}{2}(1-m^2)\frac{J^2{\cal P}_0^{*'}(m)}{{\cal P}_0^*(m)}
\end{eqnarray}
along the ascending branch.

\section{Alternative expression of $C(g,H)$}

Eqs. (\ref{Eq82}) and (\ref{Eq83}) are not  convenient for numerical calculation because  the Lambert function must be evaluated in the complex plane. This problem can be circumvented by changing the integration variable. From the equation $W(z)e^{W(z)}=z$, we have
\begin{eqnarray}
\label{Eq84}
A^{(0)}e^{-A^{(0)}}\cos 2\pi x=-e^{u}[u\cos v-v\sin v]\nonumber\\
A^{(0)}e^{-A^{(0)}}\sin 2\pi x=e^{u}[v\cos v+u\sin v]
\end{eqnarray}
where $u(x)$ and $v(x)$ are respectively the real and imaginary parts of $W(-A^{(0)}e^{-A^{(0)}}e^{-2i\pi x})$. For $W=W_0$ (resp. $W=W_{-1}$), $u$ is a continuous function of $x$ that monotonically increases (resp. decreases) as $x$ increases from $0$ to $1$ and $v$ can be univoquely expressed as a function of $u$ through the relation
\begin{eqnarray}
\label{Eq85}
v(u)=\pm(A^{(0)2}e^{-2(A^{(0)}+u)}-u^2)^{1/2} 
\end{eqnarray}
where the signs $+$ and $-$ refer to $W=W_0$ and $W=W_{-1}$, respectively. A careful analysis shows that the inverse function $x(u)$ is given by
\begin{eqnarray}
\label{Eq86}
x(u)&=\frac{1}{2\pi}\arctan \frac{u\tan v+v}{v\tan v-u} \ , \ \ -A^{(0)}\le u\le u_1 \ (\mbox{resp.} \ u_1<u\le -A^{(0)})\nonumber\\
x(u)&=\frac{1}{2}+\frac{1}{2\pi}\arctan \frac{u\tan v+v}{v\tan v-u}\ , \ \ u_1<u\le u_0\ (\mbox{resp.}\  u_0\le u\le u_1)\nonumber\\
\end{eqnarray}
where $u_0=u(x=1/2)=\Re \{W(A^{(0)}e^{-A^{(0)}})\}$ and $u_1=u(x=1/4)=\Re \{W(iA^{(0)}e^{-A^{(0)}})\}$.
Making the change of variable $x\rightarrow u$ in Eq. (\ref{Eq82}), we then obtain
\begin{eqnarray}
\label{Eq87}
\fl C(g,H)= \frac{1}{2\pi}\int_{-A^{(0)}}^{u_0}du \ [f(x(u),g)+f(1-x(u),g]\frac{[A^{(0)2}-u^2+v^2][(1+u)^2+v^2]}{v} \ ,\nonumber\\
\end{eqnarray}
where the proper expression of $x(u)$ must be used in each interval $[-A^{(0)},u_1]$ and  $]u_1,u_0]$ (resp. $[u_0,u_1]$ and  $]u_1,-A^{(0)}]$). Similarly, in Eq. (\ref{Eq83}), by using $\frac{1}{\sin^2(\pi x)}\ dx=-\frac{d}{\pi dx}\cot(\pi x)\ dx=-\frac{d}{\pi du}\cot(\pi x(u))\ du$  and integrating by parts, we get 
\begin{eqnarray}
\label{Eq88}
\frac{C(g,H)}{g}\rightarrow -\frac{2}{\pi}\int_{-A^{(0)}}^{u_0}du \frac{2u+u^2+v^2}{\tan(\pi x(u))}
\end{eqnarray}
when $g\rightarrow +\infty$. Note that the integrand in Eq. (\ref{Eq87}) diverges for both 
$u\rightarrow -A^{(0)}$ and $u\rightarrow u_0$ when $W=W_0$ (in both cases $v\rightarrow 0$), whereas it only diverges for $u\rightarrow -A^{(0)}$ when $W=W_{-1}$. However,  these are inverse square-root singularities and the integral is finite, as it must be.

\end{appendix}


\begin{thebibliography}{10}
\bibitem{note4} $m_1^{max}(H)$ as given by Eq. (\ref{Eq75b}) diverges when $A^{(0)}=2J{\cal P}_0^*\rightarrow 1$. Therefore, one must ``renormalize'' the $1/z$ expansion as explained at the end of Appendix B and consider the field $H^{max}(m)=H_0(m)-(1/z)m_1^{max}(H_0(m))[1-2J{\cal P}_0^*(H_0(m)]/2{\cal P}_0^*(H_0(m))$. The complexity $\Sigma^{max}_Q$, which only depends on $A^{(0)}$, is then considered as a function of $H^{max}$.
\end{thebibliography}
\end{document}